\documentclass[useAMS,usenatbib,fleqn]{mnras}
\usepackage{graphicx}
\usepackage{times}
\usepackage{amssymb, amsmath}
\usepackage{algpseudocode}
\usepackage{algorithm}
\usepackage[section]{placeins}
\usepackage[english]{babel}
\usepackage{lineno}
\PassOptionsToPackage{hyphens}{url}
\usepackage{ulem}
\usepackage{hyperref}
\usepackage{url}
\usepackage{diagbox}
\newcommand{\beqs}{\begin{eqnarray}}
\newcommand{\eeqs}{\end{eqnarray}}
\newcommand{\cpij}{C^P_{ij}(\theta)}
\newcommand{\hinvMpc}{h^{-1} {\rm Mpc}}      

\newcommand{\beq}{\begin{equation}}
\newcommand{\eeq}{\end{equation}}

\newcommand{\norm}[1]{\left\lVert#1\right\rVert}
\input{epsf}

\def\aap{A\&A}

\def\apj{ApJ}
\def\apjs{ApJS}
\def\apjl{ApJL}

\def\mnras{MNRAS}
\def\aj{AJ}
\def\nat{Nature}

\def\prd{PhRvD}

\def\araa{ARA\&A}

\usepackage{color}

\title[Photo-$z$ Self-calibration]{Using angular two-point correlations to
self-calibrate the photometric redshift distributions of DECaLS DR9}
\author[H. Xu et al.]{Haojie Xu$^{1,2}$\thanks{E-mail: haojie.xu@sjtu.edu.cn},
Pengjie Zhang$^{1,3,2}$,
Hui Peng$^{1,2}$,
Yu Yu$^{1,2}$,
Le Zhang$^{4,5,6}$,
Ji Yao$^{7}$,
Jian Qin$^{1,2}$,
Zeyang Sun$^{1,2}$,
\newauthor
Min He$^{1,2}$,
and Xiaohu Yang$^{1,3,2}$
\\
\\
$^1$
Department of Astronomy, Shanghai Jiao Tong University, Shanghai 200240, China\\
$^2$ Key Laboratory for Particle Astrophysics and Cosmology
(MOE)/Shanghai Key Laboratory for Particle Physics and Cosmology, China\\
$^3$
Tsung-Dao Lee Institute, Shanghai Jiao Tong University, Shanghai 200240, China\\
$^4$
School of Physics and Astronomy, Sun Yat-Sen University, Guangzhou 510297, China\\
$^5$
CSST Science Center for the Guangdong-Hong Kong-Macau Greater Bay Area, Zhuhai 519082, China\\
$^6$
Peng Cheng Laboratory, No.2, Xingke 1st Street, Shenzhen 518000, China\\
$^7$
Shanghai Astronomical Observatory (SHAO), Nandan Road 80, Shanghai 200030, China\\
}
\begin{document}
\maketitle

\begin{abstract}
Calibrating the redshift distributions of photometric galaxy samples 
is essential in weak lensing studies. The self-calibration method 
combines angular auto- and cross-correlations between galaxies 
in multiple photometric redshift (photo-$z$) bins to reconstruct 
the scattering rates matrix between redshift bins. 
In this paper, we test a recently proposed self-calibration algorithm using the DECaLS Data Release 9 and investigate to what extent the scattering rates are determined. 
We first mitigate the spurious angular correlations 
due to imaging systematics by a machine learning based method. We then improve the algorithm for $\chi^2$ minimization and error estimation.  
Finally, we solve for the scattering matrices, carry out a series of consistency tests and find reasonable agreements: (1) finer photo-$z$ bins return a high-resolution scattering matrix, and 
it is broadly consistent with the low-resolution matrix from wider bins; 
(2) the scattering matrix from the Northern Galactic Cap is almost 
identical to that from Southern Galactic Cap; (3) the scattering matrices are in reasonable agreement with those constructed from the power spectrum and the weighted spectroscopic subsample. 
We also evaluate the impact of cosmic magnification. Although it changes little the diagonal elements of the scattering matrix, it affects the off-diagonals significantly. 
The scattering matrix also shows some dependence on scale cut of input correlations, which may be related to a known numerical degeneracy 
between certain scattering pairs.
This work demonstrates the feasibility of the self-calibration method in real data
and provides a practical alternative to calibrate the redshift distributions
of photometric samples.

\end{abstract}


\section{Introduction}
\label{sec:intro}

One outstanding question for modern cosmology is to tell whether the accelerating expansion
of the Universe is due to the so-called dark energy or indicating that 
general relativity fails at cosmological scales. 
Weak gravitational lensing is a powerful probe to constrain such models
since it is sensitive to both distance--redshift relation and the time-dependent
growth of structure \citep{Albrecht2006}. However, the accuracy of 
photometric redshift casts a shadow over the inferred 
cosmological parameter confidence.
In a typical 3$\times$2-point analysis, 
the Large Synoptic Survey Telescope (LSST) Dark Energy Science Collaboration 
\citep{LSST2018} requires a $\sim0.001$ accuracy for mean redshift in tomographic bins  (e.g., \citealt{Ma2006}). More accurate
photometric redshift (photo-$z$) distribution leads to more accurate cosmological inference. 
Therefore, it is crucial to have an accurate photo-$z$ distribution at the current precision cosmology age.

There are usually consisting of two steps to obtain an accurate photo-$z$ distribution.
The first step is designing some algorithms to estimate photo-$z$
for each galaxy as accurately as possible, given a few broadband photometry. 
Such algorithms have been well studied, from the
traditional template-fitting method to machine learning based technique and hybrid 
(see a recent review in \citealt{Salvato2019}).
Given the photo-$z$ estimation for each galaxy, one can divide galaxies into a few tomographic bins.
The second step is calibrating 
the photo-$z$ distributions to minimize the mean redshift uncertainty for tomographic bins.

Depending on whether using an external reference sample with secure redshift, 
the photo-$z$ calibration methodology can be roughly divided into two categories. 
The direct calibration weights the galaxies in the reference sample to match the galaxy distribution in multidimensional colour--magnitude space \citep{Lima2008, Bonnett2016, Hildebrandt2017}. 
Along the same line, a machine learning based method, self-organizing map (SOM), has been
recently introduced to the field to better characterize the bias and 
uncertainty due to non-representative
or incompleteness of the reference sample (e.g., \citealt{Masters2015,Masters2017,Masters2019, Buchs2019, Davidzon2019, Wright2020a, Hildebrandt2021}).
If the reference sample spatially overlaps with the photometric
sample, the physical association between them due to the large-scale structure can help constrain the redshift distributions of the photometric
sample.
This clustering-based method is another major branch of utilizing a reference sample. 
By cross-correlating with the reference sample, the redshift distribution of the photometric sample can be well reconstructed (see e.g., \citealt{Newman2008, Matthews2010, Matthews2012, Schmidt2013, McQuinn2013, Menard2014, Kovetz2017, McLeod2017, Busch2020, Gatti2018, Gatti2021}). This method does not require the representativeness or completeness of the reference sample. However, the redshift-independent galaxy bias assumption inside tomographic bins adopted in this method might bias the inferred photo-$z$ distribution (e.g., \citealt{Davis2018}). 
See a recent comprehensive review \citep{Newman2022} for more details on the clustering-based method.

On the other hand, several methods can recover the redshift distribution of a photometric sample without any reference to the redshift sample. 
For instance, \citet{Quadri2010} counted close galaxy pairs in angular positions and statistically derived a measure of photo-$z$ uncertainty from
the differences in photo-$z$ of close pairs.
Instead of cross-correlating with a reference sample, the cross-correlation between photo-$z$
bins themselves can also infer the contamination fractions (or scattering rates) between redshift bins, according to the relative amplitudes of auto- and cross-correlations. The idea 
relies on the fact that contamination between bins will result in non-zero cross-correlations between those bins, whose amplitude is proportional to the contamination fractions.
\citet{Schneider2006} discussed how well angular correlations could constrain
the parameters in their photo-$z$ error model. In addition to the galaxy--galaxy clustering, \citet{Zhang2010} advocated that the shear-galaxy cross-correlations
from the same set of photo-$z$ bins should help break the severe degeneracy in contamination fractions inferred from galaxy--galaxy clustering alone (see e.g. Fig.11 in \citealt{Erben2009} and Fig.2 in \citealt{Benjamin2010}). 
However, both discussions in \citet{Schneider2006} and \citet{Zhang2010} are forecasts 
from the Fisher matrix formalism, and a practical algorithm for solving scattering
rates is still needed.
 
\citet{Erben2009} first managed to solve the scattering rates between two photo-$z$ bins. They
found their estimated results consistent with that from the direct calibration method. 
The so-called ``pairwise analysis'' is
further extended in \citet{Benjamin2010} to a multibin scenario,  
where they performed ``pairwise analysis'' to each pair of tomographic bins. 
The extension implicitly ignored the common contamination from a third redshift bin.
Their method was tested against mock catalogues and found a good recovery.
They then applied the method to observation data from the Canada–-France–-Hawaii Telescope Legacy Survey (see also \citealt{Benjamin2013}). However, the systematic bias due to the simplifications adopted in the pairwise analysis might not meet the stringent requirement of stage IV projects like LSST.
Without any simplifications, \citet{Zhang2017} proposed an accurate and efficient algorithm (self-calibration algorithm hereafter)
to overcome the constrained non-linear optimization problem encountered 
in the self-calibration method.
With incorporating the shear-galaxy cross-correlations, the algorithm demonstrated to recover contamination fractions at the accuracy level of 0.01-1 per cent, and the accuracy of mean redshift in photo-$z$ bins can reach to 0.001, i.e., $\delta_{\langle z \rangle}\sim$ 0.001.

In this work, we implement the self-calibration algorithm to 
calibrate the redshift distributions of photometric galaxies from the
DECaLS Data Release 9. 
We employ a machine learning method to 
mitigate the spurious correlations introduced by imaging systematics,
such as foreground stellar contamination, Galactic extinction, and seeing. 
For stability, we improve the self-calibration algorithm by starting from a more 
reasonable initial guess and adjusting the stop criteria for iterations.
Among iterations, we choose the best solution with $\chi^2$ and propagate the measurement errors to the final scattering rates matrix. 
We test our results by carrying out a series of consistency checks.
We find that the scattering matrices are broadly consistent with each other, which
demonstrates the feasibility of the self-calibration method. 

The paper is organized as follows. 
We describe the data set, imaging systematics mitigation, and angular correlation measurements in section~\ref{sec:data}. 
In section~\ref{sec:method} we briefly introduce the self-calibration algorithm. 
The main results and tests are presented in section~\ref{sec:results}.
Finally, we summarize and discuss our results in section~\ref{sec:dis}.
We use the terms tomographic bin and photo-$z$ bin interchangeably.

\section{Galaxy Samples, Imaging systematics, and Clustering Measurements}
\label{sec:data}

We investigate the scattering rates between redshift bins 
based on the publicly available catalogue Photometric Redshifts
estimation for the Legacy Surveys (PRLS \footnote{\url{https://www.legacysurvey.org/dr9/files/\#photometric-redshift-sweeps}}; \citealt{Zhou2020}). 
The PRLS catalogue is constructed from the photometry of the Data Release 9 of the DESI Legacy Imaging Surveys (LS DR9).
The optical imaging data ($g,r,z$ bands) of LS DR9 is contributed by three projects: the Beijing--Arizona Sky Survey (BASS) for $g,r$ bands in the North Galactic Cap (NGC) at declinations $\delta \geqslant 32.375^\circ$, the Mayall z-band legacy Survey (MzLS) for $z$ band in the same footprint as BASS, and the Dark Energy Camera Legacy Survey (DECaLS) for $g,r,z$ bands in the North Galactic Cap (DECaLS-NGC) at $\delta \leq 32.375^\circ$ and the entire South Galactic Cap (DECaLS-SGC, including the data contributed by Dark Energy Survey). These three surveys covers $\sim 18000$ deg$^2$ sky area in total, among which the BASS/MzLS covers $\sim 5000$ and $\sim 13000$ deg$^2$ from the DECaLS. The photo-$z$ estimation also utilized the two mid-infrared (W1 3.4 $\mu$m, W2 4.6 $\mu$m) imaging from the Wide-field Infrared Survey Explorer (WISE) satellite. 
An overview of the surveys can be found in \citet{Dey2019}. 

\begin{figure}
\includegraphics[width = \columnwidth]{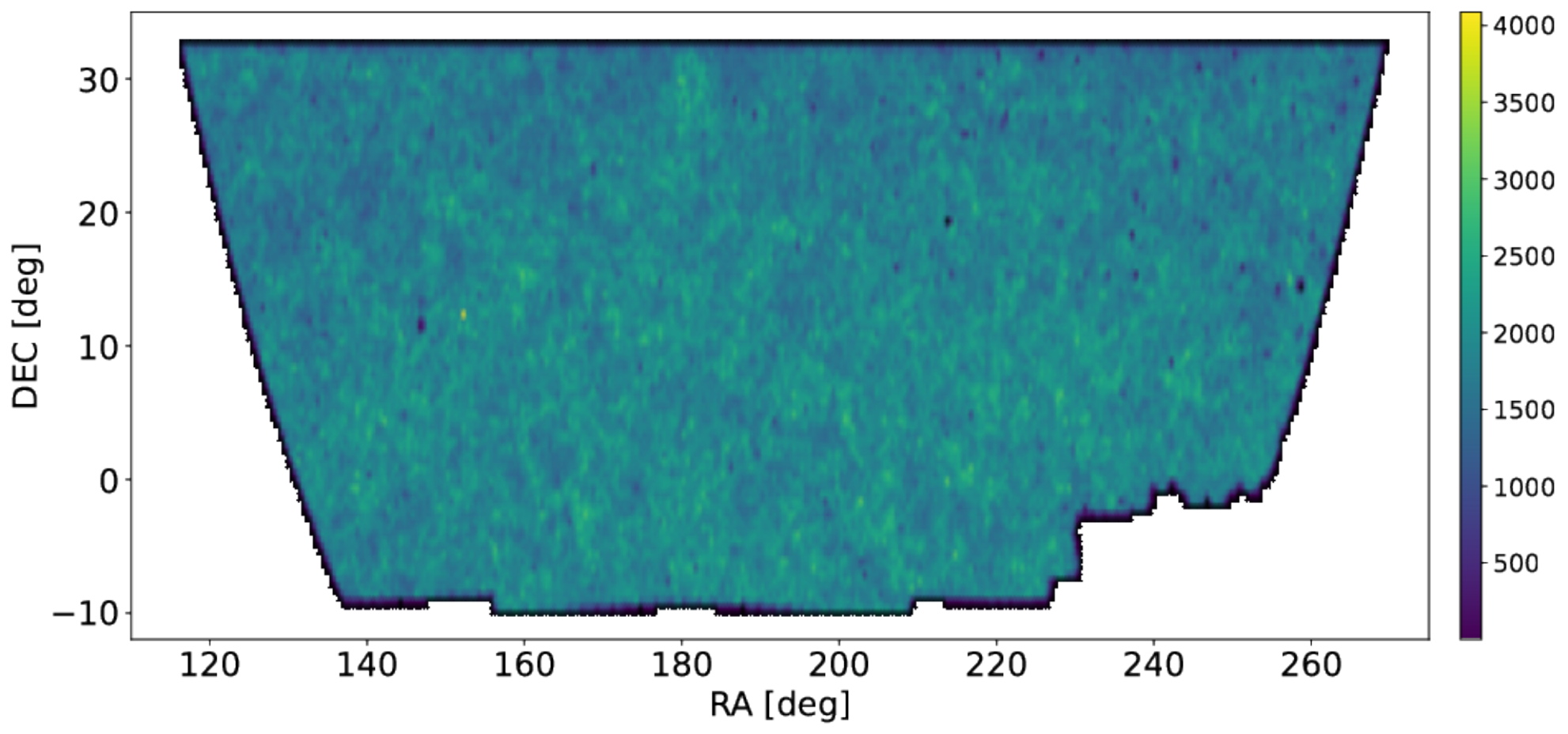}	
\caption{
The footprint of the default sample used in this work. 
The galaxy population is $\sim 40$ million with a sky
coverage $\sim 5230$ deg$^2$, with a surface density $\sim 2.13$ galaxies
per square arcmin. The colour code is the galaxy number counts per 0.25 deg$^2$. Note that, we do not take into account the  
fractional observed area (section~\ref{subsec:imaging}) in this plot. 
}
\label{fig:footprint}
\end{figure}

Since the primary purpose of this work is to apply the self-calibration algorithm to observational photometric survey, we present our fiducial results on the DECaLS-NGC region, whose footprint (after selection in section~\ref{subsec:sample}) is shown in Fig.~\ref{fig:footprint}. We bin the DECaLS-NGC galaxies into 5 equal width tomography bins 
in $0<z_{\rm photo}<1$ as our fiducial analysis. Note that we adopt the median value of photo-$z$ reported in PRLS as our default $z_{\rm photo}$. 

\subsection{Sample selections and photo-$z$ estimation}
\label{subsec:sample}
We construct our galaxy samples for clustering measurements 
similar to the selection criterion in section 2.1 of \citet{Yang2021}.
We briefly summarize the steps here.
First, we select out ``galaxies'' (extended imaging objects) according to the 
morphological types provided by the {\sc tractor} software \citep{Lang2016}. 
To have a reliable photo-$z$ estimation, we select 
objects with at least one exposure in each optical band.
We also remove the objects 
within ${\rm |b| < 25.0^{\circ}}$ (where ${\rm b}$ is the Galactic latitude) 
to avoid high stellar density regions. Finally, we remove any objects
whose fluxes are affected by the bright stars, large galaxies, or globular
clusters (maskbits 1, 5, 6, 7, 8, 9, 11, 12, 13 \footnote{\url{https://www.legacysurvey.org/dr9/bitmasks/}}). 
The final sky coverage of our galaxy sample is shown in Fig.~\ref{fig:footprint}.
We apply identical selections to the publicly available random catalogues \footnote{\url{https://www.legacysurvey.org/dr9/files/\#random-catalogues-randoms}}.

\begin{figure}
\includegraphics[width = \columnwidth]{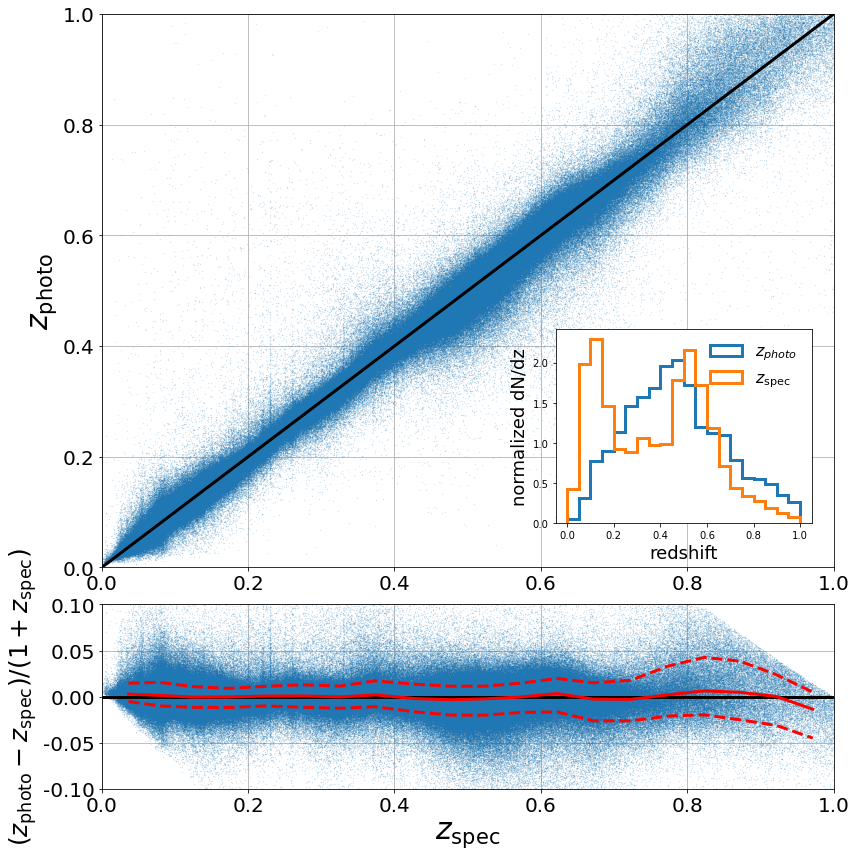}	
\caption{
Photometric redshift performance of the full sample.
{\it Top}: The $z_{\rm photo}$ versus $z_{\rm spec}$ plot 
for the default sample. 
There are $\sim 0.92$ million out of 40 million galaxies with accurate redshift, 
marked as blue dots. The solid black line is the 1:1 line. 
The insert panel compares the normalized redshift distribution 
for spectroscopic subsample and the entire photometric sample.
{\it Bottom}: The fractional difference between $z_{\rm photo}$ and $z_{\rm spec}$ as a 
function of $z_{\rm spec}$. The solid (dashed) red line is the median (68 percentile).
The $\sigma_{\rm NMAD}$ 
and outlier rate are $\sim 0.013$ and $\sim 0.7$\%,
respectively.
Note that the performance shown here is for the full sample, which is without
conditioning on a tomographic binning. Performance tends to be worse after 
splitting galaxies into tomographic bins, 
as the training or priors are usually not conditioned on binning.
Therefore, one should not compare this figure with the performance evaluated in tomographic bins, e.g., Fig.~\ref{fig:pij_5_redshift_bins}.
}
\label{fig:specz_photoz}
\end{figure}

The photometric redshift provided in the PRLS catalogue is inferred by
the random forest method, a machine learning algorithm
that takes advantage of the existing 
abundant galaxies with spectroscopic redshift. It
easily takes into account non-photometry attributes 
(for example, galaxy shapes) that could more significantly
improve the accuracy of photo-$z$ estimation than the traditional template-fitting 
method. Since galaxy properties (e.g., colours, magnitudes, and morphology) 
correlate with its redshift, one can predict the redshift from those properties. 
\citealt{Zhou2020} performed a regression on 
the following eight photometric parameters to the available secure 
redshift: $r$-band magnitude, $g-r$, $r-z$, $z-W1$,
$W1-W2$, half-light radius, axis ratio, and shape probability. 
The secure redshift
is adopted from various spectroscopic surveys and the COSMOS. 
The total number of galaxies with accurate redshift cross-matched with the PRLS sources 
is $\sim 1.5$ million, but only $\sim 0.67$ million of them are used
as training sample to avoid biasing the photo-$z$ estimation caused by
non-uniform distribution in the multidimensional colour--magnitude space.
The insert panel in Fig.~\ref{fig:specz_photoz} highlights the difference in redshift 
distribution of the spectroscopic subsample and the entire photometric sample.
The photo-$z$ in the PRLS catalogue
is well tested for luminous red galaxy (LRG) sample \citep{Zhou2020}. 
Besides the LRG sample, they also
demonstrated that the photo-$z$ performance for galaxies with $z_{\rm mag}<21$ 
is reasonably good, reaching a $\sigma_{\rm NMAD} \sim 0.013$
and outlier rate $\sim 1.51$\% (see their Figure B1 and B2). 
The $\sigma_{\rm NMAD}$ measures how close estimated photo-$z$ are to 
their accurate redshift, defined as 
$\sigma_{\rm NMAD} = 1.48\times{\rm median}(|z_{\rm photo}-z_{\rm spec}|)/(1+z_{\rm spec})$. 
The outlier rate is defined $|z_{\rm photo}-z_{\rm spec}| > 0.1\times(1+z_{\rm spec})$. 
We choose to apply the self-calibration algorithm to galaxy samples 
with relatively reliable photo-$z$, i.e. 
galaxies with $z_{\rm mag}<21$. 
Combined with angular masks, it ends up with $\sim 40$ million galaxies 
for the clustering measurements and $\sim 0.92$ million among them 
having accurate redshift. 
The photo-$z$ performance of the full sample is 
shown in Fig.~\ref{fig:specz_photoz}. The $\sigma_{\rm NMAD}$ 
and outlier rate for our default sample are $\sim 0.013$ and $\sim 0.7$\%,
respectively. Note that, because the spectroscopic subsample 
is relatively brighter, the good performance shown in Fig.~\ref{fig:specz_photoz}
can not be naively extrapolated to the whole sample. 
However, one can still expect
that the photo-$z$ performance for the whole sample would not be too far
away from that of the spectroscopic subsample.

\begin{figure*}
\includegraphics[width=\textwidth]{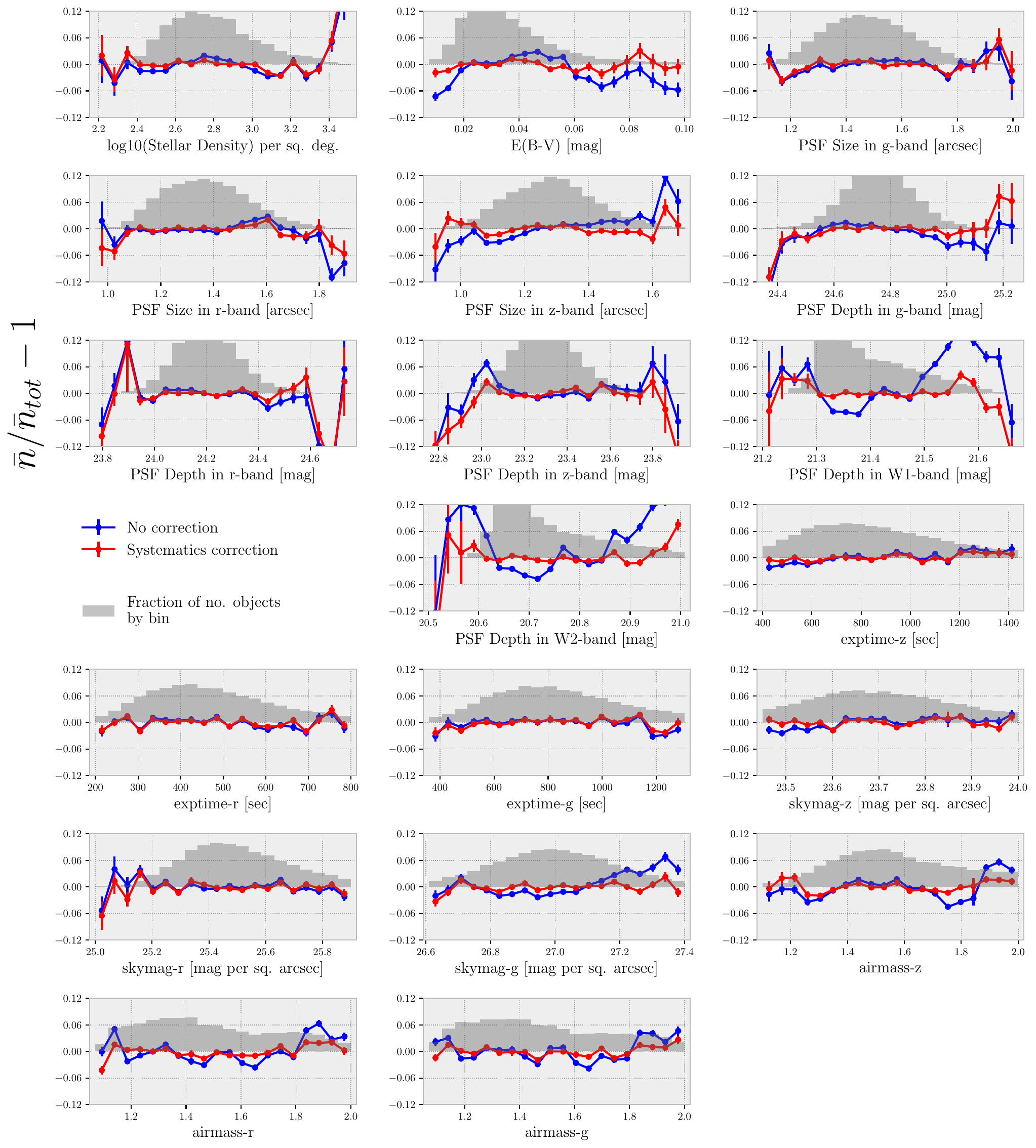}
\caption{
Fractional galaxy overdensity as a function of 19 input imaging maps.
The blue/red lines show the relative density before/after the systematics 
correction inside those valid pixels. 
The histogram is the fraction of pixels (or galaxies) in bins of imaging 
properties, which is used to estimate the errors as the standard deviation
of galaxies in bins. 
After mitigation, 
the weighted galaxy densities show little fluctuation with respective to 
imaging properties.
Since imaging systematics may depend on sample's luminosity and redshift,
we apply the imaging systematics mitigation procedure individually 
to all interested samples before measuring the correlations. 
To avoid clutter, we here showcase 
the first tomographic sample of $0< z < 0.2$. 
Plots for other samples are available on request.
}
\label{fig:systematics}
\end{figure*}

\subsection{Imaging systematics mitigation}
\label{subsec:imaging}
Observation conditions, such as stellar contamination, Galactic 
extinction, sky brightness, seeing, and airmass, 
introduce spurious fluctuations in the observed galaxy density
\citep{Scranton2002a, Myers2006, Ross2011a, Ross2017, Ho2012, Morrison2015}.
Therefore, any direct clustering measurements from photometric samples
are potentially biased, especially on large scales.
Three approaches have been taken to mitigate the imaging systematics. 
The first approach assumes that the observed overdensity
field is the sum of true cosmological overdensity and 
some function of over-densities of various imaging maps.
The functional form can be complicated. 
A weighted linear relation is firstly employed in the early studies 
and one can work out 
the weights for imaging maps using the auto- and cross-correlations between
imaging maps and observed galaxy density \citep{Ross2011a, Ho2012}. 

Rather than working out weights by the correlations, an alternative is
to translate it to a linear regression problem: various imaging
maps being the independent variables (features) and observed 
overdensity being the dependent variable (label). 
One can linearly fit the impact of imaging systematics on observed overdensity.
The best-fitting weights can be achieved by minimizing the difference between
predicted and observed target overdensity \citep{Myers2015, Prakash2016}. 
Note that two sets of weights are different: 
In the clustering scenario, the weights are mapwise,
i.e., one imaging map shares one weight; 
while in the regression scenario, the weights are pixelwise,
i.e., pixels in the same imaging space share the same weights. 
The weights from
regression approach can be directly applied to the observed and random galaxies
to mitigate the imaging systematics.

However, the linear assumption may fail to capture the non-linear
dependence on imaging systematics in some strong contamination regions, 
for example, close to Galactic plane (see e.g., \citealt{Ho2012}). 
A recent progress is to relax
the linear dependence assumption to more flexible functions. The 
regression approach has the merit that it can be easily generalized to flexible functions by 
machine learning algorithms, such as Random Forest (RF) and Neural Network (NN).

The other two approaches that help remove the imaging
systematics are the mode-projection based technique 
\citep{Tegmark1998,Leistedt2013,Elsner2016,Kalus2019} and the 
forward-modelling approach \citep{Berge2013, Suchyta2016, Kong2020}. 
In this work, we adopt the machine learning based regression approach.

\begin{figure}
\includegraphics[width = \columnwidth]{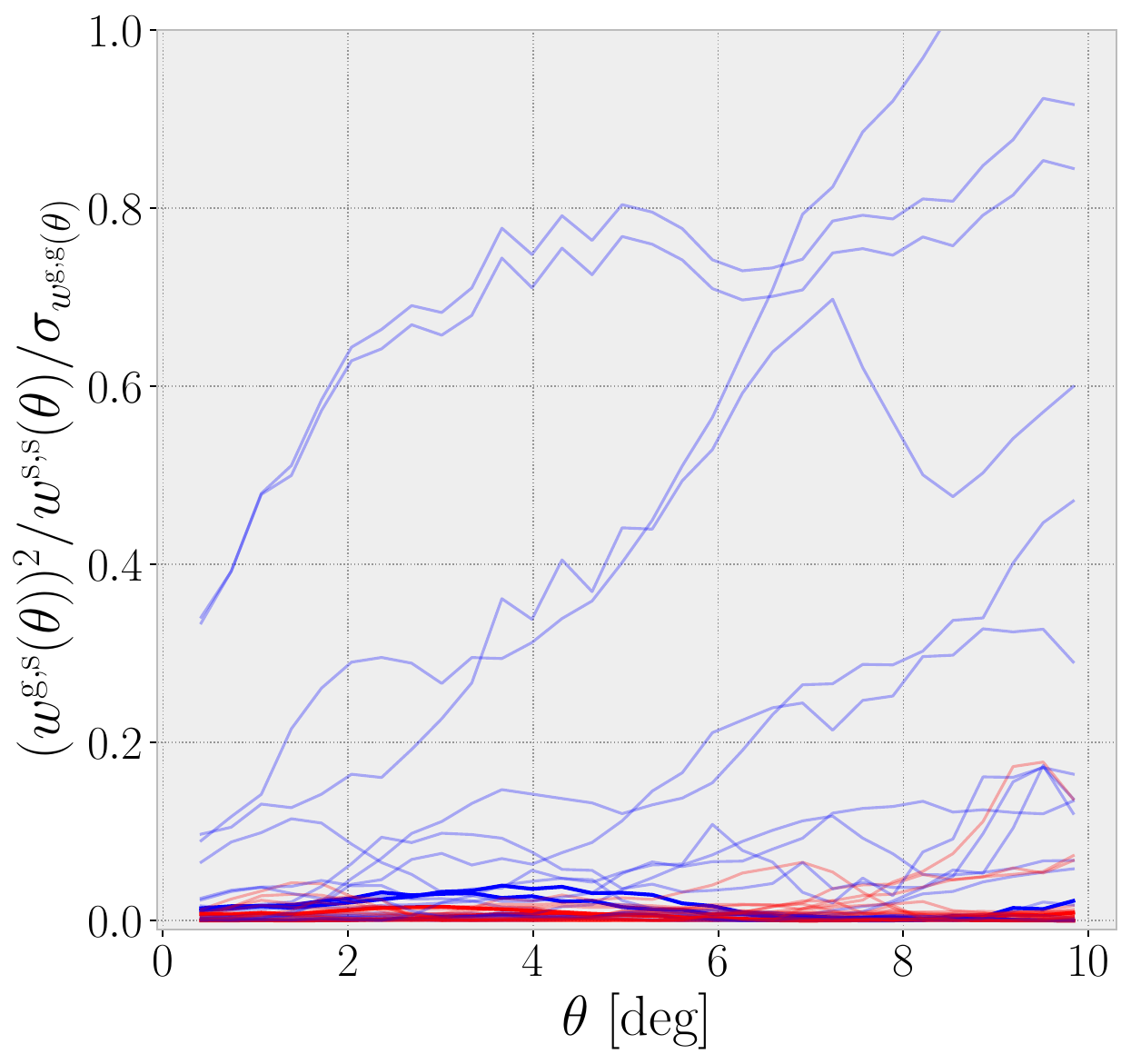}	
\caption{
The imaging contamination in galaxy autocorrelations before the
mitigation (blue lines) and after the mitigation (red lines), normalized to the
errors of galaxy autocorrelations before the mitigation. 
After mitigation, the systematic bias induced by various imaging systematics
relative to the statistical uncertainty is approaching zero, significantly 
lower than that before mitigation. There are in total 21 pairs of blue and red 
lines shown in the plot. Besides the 19 input training imaging maps (faint lines), 
we include two additional maps (bold lines), the Sagittarius Stream stellar maps and 
Hydrogen atom column density. The result that two red bold lines are closer to zero than their blue counterparts reflect the redundancy of imaging maps, 
though this tomographic sample seems to suffer little from these two test imaging properties. 
Again, we showcase the first tomographic sample of $0< z < 0.2$ for illustration, and all other samples display similar trends. 
}
\label{fig:systematics_cross}
\end{figure}

\begin{figure*}
\includegraphics[width=\textwidth]{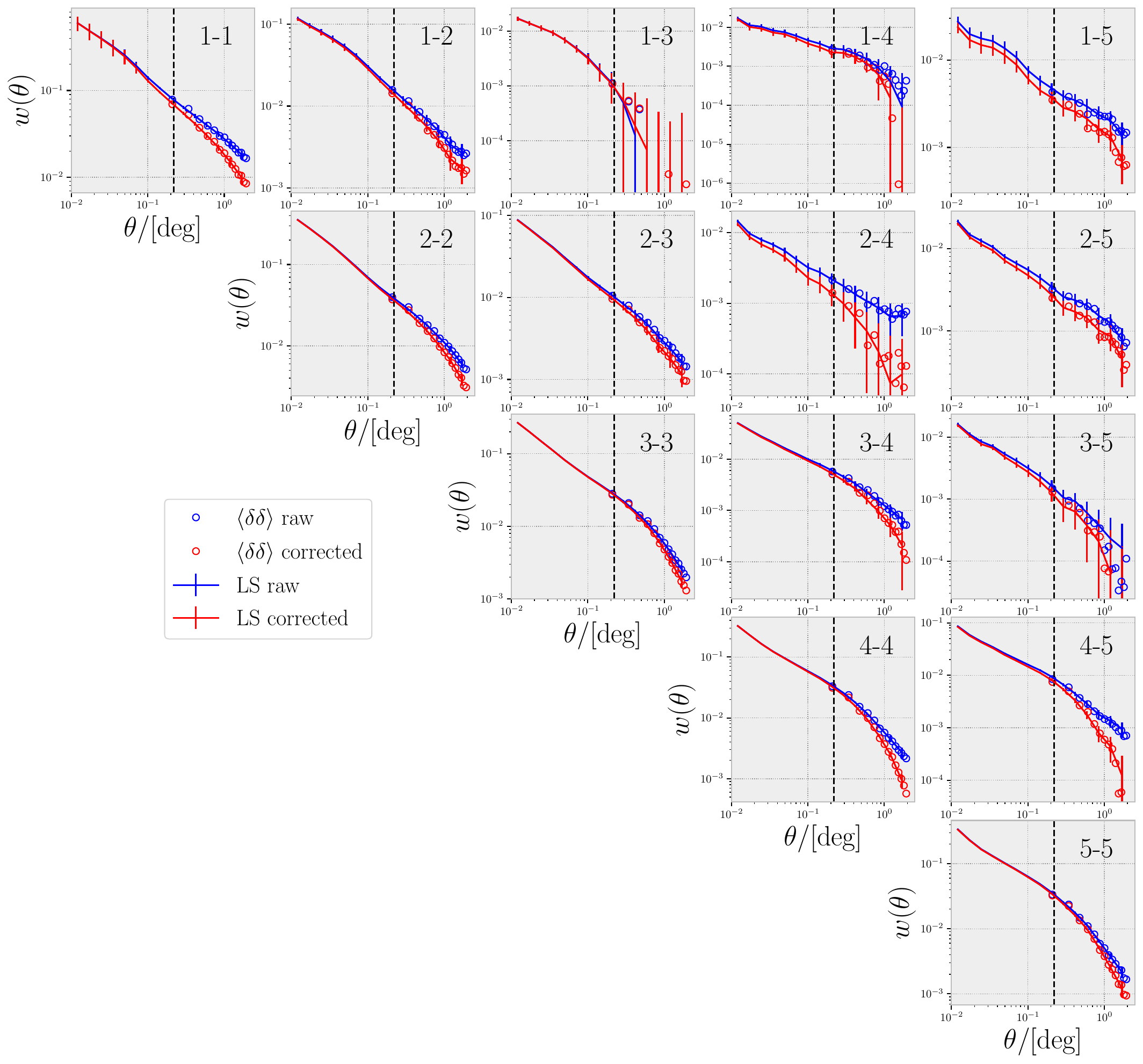}
\caption{
The auto- and cross-
angular correlation function before and after the mitigation
for five tomographic bins. The tomographic bin indices are indicated
at the top right corner in each panel, where indices 1, 2, 3, 4, and 5 
represent the 
photo-$z$ bin (0, 0.2), (0.2, 0.4), (0.4, 0.6), (0.6, 0.8), and (0.8, 1.0), respectively.
The lines are measured from the Landy--Szalay estimator (cf. Eq.\ref{eq:LS}) 
while the circles are from the \textsc{healpix}-based estimator (cf. Eq.\ref{eq:healpix}).
The blue (red) lines and circles represent the measurements with (without) imaging
systematics. The vertical dashed lines mark the pixel scale ($\sim 0.23$ deg)
above which the imaging systematics is largely removed. 
The error bars are estimated via the jackknife resampling method on the 
Landy--Szalay estimator. Note that, the scales of $y$-axis are different
across panels. 
}
\label{fig:2pcf}
\end{figure*}

A few studies have investigated the imaging systematics
impact on the clustering measurements in the earlier data releases of the DESI
Legacy Imaging Surveys. For Data Release 7, \citet{Kitanidis2019} studied
to what extent imaging systematics could impact clustering measurements of
various DESI target classes.
\citet{Rezaie2020} investigated the imaging systematics in Emission Line
Galaxies (ELGs) and proposed a powerful NN approach to remove the non-linear effects. \citet{Zarrouk2021} further applied the NN method to the ELGs in Data Release 8 and cross-correlated them with eBOSS QSOs to investigate
baryon acoustic oscillations. \citet{Chaussidon2021} studied the impact of imaging systematics on QSOs of Data Release 9 and compared the linear, RF, and NN regression on mitigation effect. 

Here, we choose to apply the Random Forest mitigation technique of \citet{Chaussidon2021} 
to our galaxy samples for two reasons: 1, 
RF performs better than the linear or quadratic method; 
2, compared to NN, RF reaches similar results but with less
computational expense. 
We briefly summarize the main steps here, and a detailed description
of the methodology is provided in \citet{Chaussidon2021}. We also note that 
most of the following procedures are handily encapsulated in the GitHub code
{\sc regressis}\footnote{\url{https://github.com/echaussidon/regressis}}.

\begin{itemize}
\item (1) The imaging maps adopted in this work are kindly provided by
E. Chaussidon (private communication). These maps are generated from code 
script \textit{bin/make\_imaging\_weight\_map} from the {\sc desitarget}
\footnote{\url{https://github.com/desihub/desitarget}} package with 
{\sc healpix}\citep{Healpix} $N_{\rm side}=256$ (resolution of $\sim 0.23$ deg).
We select the following photometric properties as our imaging features:
stellar density \citep{Gaia2018}, Galactic extinction\citep{Schlegel1998,Schlafly2011}, sky brightness 
(in $g/r/z$ bands), airmass ($g/r/z$), exposure time ($g/r/z$), PSF size ($g/r/z$), and PSF depth ($g/r/z$/W1/W2). In total, we have 19 imaging maps. The galaxy density maps are binned of the same resolution.
\item (2) There are 99563 pixels inside our default sample (cf. Fig.\ref{fig:footprint}). Among them, we only consider pixels with $f_{{\rm pix}, i} > 0.8$,
where $f_{{\rm pix}, i}$ is the fractional observed area of pixel $i$ calculated as the ratio of random points after the selection (section~\ref{subsec:sample}) and before the selection. This cut makes sure our regression is only performed on reliable pixels.
Combined with some pixels with NAN imaging properties, we trim out in total 
2197 pixels, accounting for 2.21 per cent of our sample footprint.
\item (3) To avoid overfitting, we divide the sample into $K$ folds, and the training
is only performed on $K$-1 folds and the rest one fold for prediction. 
As suggested in \citet{Chaussidon2021}, $K=7$ is adopted for the default sample, making each fold cover $\sim 830$ deg$^2$. The number of folds makes sure the regression is efficient and less
prone to overfitting. Since the imaging systematics is probably region dependent, 
the locations of folds matter. The algorithm cannot predict the Galactic contamination if all the training folds are located at high Galactic latitudes. Therefore, each fold should contain pixels across the entire footprint. The goal is achieved by applying the GroupKFold function from {\sc SCIKIT-LEARN} package to the data set. It groups pixels into small patches that finally make up the folds. Each patch in our setting contains $\sim$1000 pixels and covers around 52 deg$^2$.
\item(4) We use the same RF hyperparameters as in \citet{Chaussidon2021}, i.e., 200 decision trees, the 
minimum number of samples (pixels in this scenario) at a leaf node being 20. 
We tweak these hyperparameters and find no significant difference in systematics reduction. 
\end{itemize}

After these steps, a weight factor $w_{{\rm pix}, i}$ for each 
valid pixel $i$ will be returned, and the 
imaging systematics will be reduced by weighting the galaxies according to their pixel weights.
Galaxies in the same pixel share the same weight and therefore, imaging systematics is 
removed above the pixel scale. 
We apply the mitigation procedures individually to interested tomographic samples.
To avoid cluttering, we showcase the mitigation results for 
$0 < z < 0.2$ sample. Figure~\ref{fig:systematics} shows the
galaxy overdensity, before and after the mitigation, 
as a function of 19 input imaging properties. If the galaxy
density field is independent of imaging properties, one would 
expect the mean galaxy density in bins of imaging properties
amounts to the global mean. However, it is clear that the
galaxy sample suffers some imaging contamination 
(blue lines in Fig.~\ref{fig:systematics}). 
After the mitigation, the corrected density (red lines 
in Fig.~\ref{fig:systematics}) is almost flat for all imaging
properties, though some large fluctuations are expected at margins.

We further compare the cross-correlations between 
galaxy densities, before and after mitigation, 
and various imaging maps, as shown in
Fig.~\ref{fig:systematics_cross}. 
We adopt the {\sc healpix}-based estimator to 
calculate the angular correlation function, where for a separation angle
$\theta$, is defined as (see e.g., \citealt{Scranton2002a, Ross2011a, Rezaie2020})
\begin{equation}
\omega^{p, q}(\theta)=\frac{\sum_{i j} \delta_{{\rm pix},i}^{p} \delta_{{\rm pix}, j}^{q} \Theta_{i j}(\theta) f_{\mathrm{pix}, i} f_{\mathrm{pix}, j}}{\sum_{i j} \Theta_{i j}(\theta) f_{\mathrm{pix}, i} f_{\mathrm{pix}, j}},
\label{eq:healpix}
\end{equation}
where $p = q$ gives the autocorrelation function estimator, and
$p \neq q$ gives the cross-correlation function estimator.
$\delta^{p}_{{\rm pix}, i}$ represents the overdensity in pixel $i$
of field $p$ (e.g., galaxy density field before or after the mitigation, imaging maps), taking into account the fractional observed area by
$\delta^{p}_{{\rm pix}, i}=n^{p}_{{\rm pix}, i} /\left(f_{\mathrm{pix}, i} \bar{n}^{p}\right)-1$.
$\Theta_{i j}$ is unity when the separation angle of 
pixel $i$ and $j$ is within $\theta$ to $\theta + \Delta\theta$ 
and zero otherwise. The weight factor $f_{\mathrm{pix}, i}$
yields a higher signal to noise clustering measurements
$\omega^{p, q}(\theta)$ by giving more weight to the
pixels with higher fractional observation area.
To quantify the contamination from imaging
maps $s_{k}$ in galaxy autocorrelation $\omega^{g, g}(\theta)$, we calculate the quantity
$\left[\omega^{g, s_{k}}(\theta)\right]^{2} / \omega^{s_{k}, s_{k}}(\theta)$
\footnote{This expression gives exactly the amount of contamination in the 
galaxy autocorrelation if the observed overdensity is the sum of cosmological
overdensity and a linear combination of imaging systematics, $s_k$. }
and normalize to the error of galaxy autocorrelation
$\sigma_{\omega^{g, g}}$. The normalization highlights the systematics
bias compared to statistical uncertainty. It is clear that
after the mitigation, the corrected observed field has minimal
correlation with various imaging maps. All {\sc healpix}-based 
angular correlation functions are calculated via the 
python package {\sc treecorr}\footnote{\url{https://github.com/rmjarvis/TreeCorr}}.

\subsection{Galaxy clustering measurements}
\label{subsec:clustering}
We bin our fiducial samples into 5 and 10 equal width tomographic bins 
in $0<z_{\rm photo}<1$ and 
estimate the auto and cross angular two-point correlation functions (2PCFs) 
of the galaxies using the Landy--Szalay estimator \citep{Landy93}.
The pair counts are obtained through python package {\sc corrfunc} \citep{Sinha2019,Sinha2020}. 
The observed auto angular 2PCFs $C^P_{ij}(\theta)$
between $i$th and $j$th redshift bin are defined as
\begin{equation}
C^P_{ij}(\theta)=\frac{aD_iD_j(\theta)-bD_iR_j(\theta)-cD_jR_i(\theta)+R_iR_j(\theta)}{R_iR_j(\theta)},
\end{equation}
where $i = j$ ($i \neq j$) gives auto- (cross-) correlations and 
$DD$/$DR$/$RR$ are respectively the weighted number of 
galaxy--galaxy/galaxy–random/random-random galaxy pairs within angular separation bin $\theta \pm \Delta\theta/2$. The normalization terms are defined as
\begin{equation}
a=\frac{\sum\limits_{i \neq j} w_{i}^{R} w_{j}^{R}}{\sum\limits_{i \neq j} w_{i}^{D} w_{j}^{D}} \quad b=\frac{\sum\limits_{i \neq j} w_{i}^{R} w_{j}^{R}}{\sum\limits_{i} w_{i}^{D} \sum\limits_{j} w_{j}^{R}} \quad 
c=\frac{\sum\limits_{i \neq j} w_{i}^{R} w_{j}^{R}}{\sum\limits_{j} w_{j}^{D} \sum\limits_{i} w_{i}^{R}},
\label{eq:LS}
\end{equation}
where $w^{D}$ is imaging correction weight (cf. Section \ref{subsec:imaging}) for data and $w^{R} = 1$.
We choose 15 bins of angle separation in logarithmic space between $[0.01, 2]$ deg.
Note that, we have applied the identical footprint cuts, number of exposure times
restrictions, and maskbits to the random catalogues 
as in galaxy sample construction (cf. section~\ref{subsec:sample}).

We estimate the covariance matrix through the jackknife resampling method. 
The footprint of the galaxy sample is divided into $N_{\rm jkf}$=120 spatially contiguous and equal area  sub-regions. We measure angular 2PCFs 120 times by leaving out one different sub-region at each time, and the covariance is calculated as 119 times the variance of the 120 measurements.

Figure.~\ref{fig:2pcf} showcases the difference in measured correlation functions before
and after imaging systematics mitigation. The correction is substantial at large scales in
both auto- and cross-correlations. As a sanity check, we compare the correlations measured from the {\sc healpix}-based estimator (cf. Eq.\ref{eq:healpix}) above the imaging correction scale ($\sim 0.23$ deg) and find a good agreement. Note that the imaging systematics can be corrected to a smaller scale as long as both the imaging maps and galaxy samples are pixelized at a higher resolution (e.g., $N_{\rm side}= 512, 1024$).
In the rest of the paper, we use correlation functions measured from the Landy--Szalay estimator 
since it provides the measurements at smaller angles.

\section{Self-calibration of photometric redshift scattering rates}
\label{sec:method}

\begin{figure*}
\includegraphics[width=\textwidth]{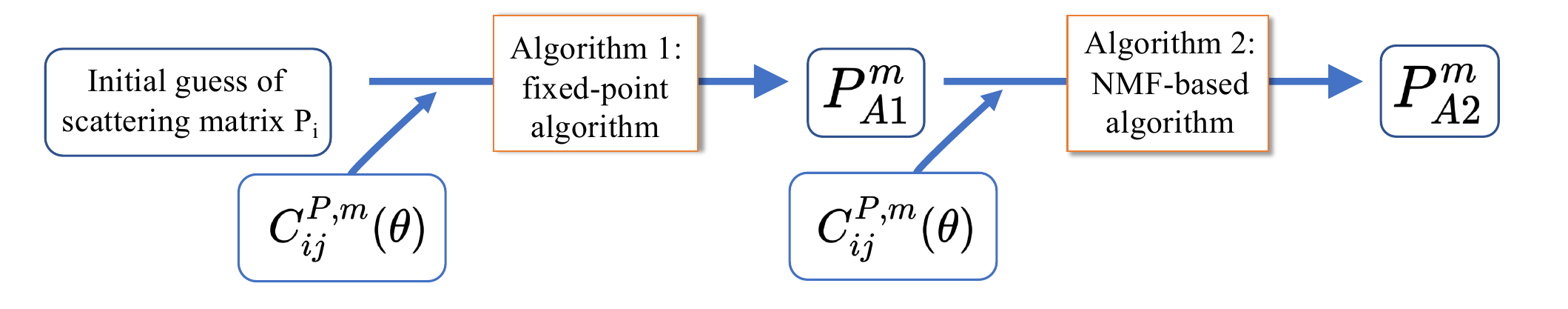}
\caption{
The flowchart of the self-calibration algorithm. 
The self-calibration algorithm is composed of two children algorithms: the fixed-point algorithm (algorithm 1) and the NMF-based algorithm (algorithm 2). 
Both algorithms iteratively search for combinations of scattering matrix $P$ and its corresponding intrinsic auto angular correlations $C^R(\theta)$ to minimize the objective $\cal{J}$ (cf. Eq.~\ref{eq:minJ}). The input to the algorithm 1 are $P_{i}$, a random diagonal-dominant matrix, and $C^{P,m}_{ij}(\theta)$, a realization of angular 2PCFs measurements $C^{P}_{ij}(\theta)$. For the algorithm 2 input, we replace the initial guess $P_{i}$ with the ``best'' scattering matrix from algorithm 1, $P^m_{A1}$, whose $\cal{J}$ value is the minimum among algorithm 1 iterations. 
The ``best'' scattering matrix from algorithm 2, $P^m_{A2}$, is defined
as the one whose $\chi^2$ (cf. Eq.~\ref{eq:chi2}) value is the minimum among algorithm 2 iterations. We treat $P^m_{A2}$ as the ``best'' scattering matrix for the realization $C^{P,m}_{ij}(\theta)$.
The median scattering matrix and its scatter are evaluated from 100 matrices of $P^m_{A2}$.
See section~\ref{sec:method} for details.
}
\label{fig:flowchart}
\end{figure*}
The differences between estimated photo-$z$ and their true-$z$ 
scatter galaxies from $i$th true-$z$ bin to
$j$th photo-$z$ bin, which introduces non-zero cross-correlations 
between photo-$z$ bins. The correlations can therefore
constrain the photo-$z$ errors \citep{Schneider2006}.
Following the convention in \citet{Zhang2010, Zhang2017}, 
the measured angular 2PCFs $\cpij$ between the $i$-th and $j$-th photo-$z$ bin can be expressed as,
\begin{equation}
\label{eq:Cgg}
\cpij=\sum_{k}P_{ki}P_{kj}C^R_{kk}(\theta)=P^TC^R(\theta)P,
\end{equation}
where $P_{ij}$ is the scattering rates from the $i$th true-$z$ bin
to $j$th photo-$z$ bin, i.e., $P_{ij} \equiv N_{i\to j}/N^P_{j}$, and $N^P_{j}$
is galaxy number counts in the $j$th photo-$z$ bin and among them, 
$N_{i\to j}$ denotes the number of galaxies from $i$th true-$z$ bin.
The second equality is written in a matrix way.
By definition, we have $\sum_i P_{ij} = 1$ and $P_{ij} \geqslant 0$.
Note that, the scattering rates matrix $P_{ij}$ is not necessary symmetric, i.e.,
$P_{ij} \ne P_{ji}$. $\cpij$ represents the measured
angular 2PCFs between $i$th and $j$th photo-$z$ bin at angular 
bin $\theta$ while $C^R_{kk}(\theta)$ is the corresponding intrinsic angular 2PCFs in $k$th true-$z$ bin. Given a small angle $\theta_k$, 
the matrix $C^R_{ij}(\theta_k)$ 
is assumed to be diagonal, i.e., $C^R_{i\ne j}(\theta_k)=0$, as the 
intrinsic cross-correlations vanish under the Limber approximation.
This assumption should be reasonable as long as the photo-$z$ bin width is not
too narrow, i.e., $\Delta z > 0.1$, whose comoving separation 
$\sim 300~\hinvMpc$ is much larger than the BAO scale.

Since Eq.~\ref{eq:Cgg} is a list of equations, one can deterministically
solve for scattering rates matrix $P$ if there are more {\it independent}  
equations than unknown variables. For example, in the configuration of
$N_z$ photo-$z$ bins and $N_\theta$ angular bins, one may expect that
there are $N_\theta N_z(N_z+1)/2$
\footnote{The factor of 2 in the denominator is simply due to
$C^P_{ij}(\theta) = C^P_{ji}(\theta)$} knowns from measured angular 2PCFs, $N_z(N_z-1)$ unknowns in scattering matrix, and $N_zN_\theta$ unknowns in intrinsic 2PCFs. 
If all the equations in Eq.~\ref{eq:Cgg} are independent, 
$N_\theta \geqslant 2$ would be enough to solve for $P_{ij}$ and
$C^R_{ii}$. In the context of cosmology, the independence of Eq.~\ref{eq:Cgg} 
is determined by the self-similarity of intrinsic
galaxy 2PCFs $C^R_{ii}$, i.e., these equations will be linear dependent if
$C^R_{ii}(\theta) \propto \theta^{\alpha}$, when the power-law index
$\alpha$ is assumed to be independent of redshift. 
In reality, several factors may affect solving
Eq.~\ref{eq:Cgg}. The first factor is the correlation between angular 2PCFs $\cpij$. 
The angular 2PCFs
$\cpij$ are usually highly correlated, where the strong correlations
on both small and large scales reflect the nature of galaxy
occupation in their parent haloes. 
The strong correlations corrupt the independence of measurements,
effectively lowering the constraining power of the self-calibration method. 
Unless performing some PCA or SVD analysis, it is otherwise
hard to tell how many measurements are independent. 
Fortunately, the intrinsic galaxy 2PCFs depart from 
a power law \citep{Zehavi04}, especially at small scales and for bright samples.
Therefore, our precise measurements at small scales would help to break the
degeneracy of the scattering matrix $P$.

\citet{Erben2009} solved Eq.~\ref{eq:Cgg} analytically for two photo-$z$ bins and 
\citet{Benjamin2010, Benjamin2013} extended their methodology to multiple bins by ignoring the common contamination in  $i$ and $j$th photo-$z$ bin from $k$th true-$z$ bin, i.e., the third term in the following equation $C_{i \neq j}^{P}(\theta)=P_{i \rightarrow i} P_{i \rightarrow j} C_{i}^{R}(\theta)+P_{j \rightarrow i} P_{j \rightarrow j} C_{j}^{R}(\theta)$
$+\sum_{k \neq i, k \neq j} P_{k \rightarrow i} P_{k \rightarrow j} C_{k}^{R}(\theta)$.
Such simplification may lead to a biased photo-$z$ calibration. 
Without any simplification, it is challenging to solve the equations of Eq.~\ref{eq:Cgg} 
given their quadratic dependence on $P_{ij}$ and linear dependence on $C^R_{ii}$. 
At the same time, the following constraints must be satisfied
when solving Eq.~\ref{eq:Cgg}, 
\begin{align}
\label{eq:constraints}
&P_{ij},\ C^{R}_{ii}(\theta) \geq 0\,, \text{\,for all~}\, i,j,\theta  \nonumber \\
&\sum_i P_{ij}=1\,, \text{\,for all~}\, j  \\
&C^{R}_{ii}(\theta) = \mathrm{Diag}\{C^{R}_{ii}(\theta)\},\, \text{\,for all~}\,  \nonumber i \nonumber 
\end{align}
\citet{Zhang2017} (Zhang17 hereafter) proposed a novel algorithm 
(self-calibration algorithm hereafter) to  numerically solve
for Eq.\ref{eq:Cgg}.
The self-calibration algorithm can be broken 
into two steps: the first step 
adopts the fixed-point method to solve 
Eq.\ref{eq:Cgg} with a fast convergence rate. 
It also has the advantage of being not 
trapped by the local minimum. 
With perfect measurements of $\cpij$,
the fixed-point algorithm should be good enough to obtain the solutions.
However, the real-world measurements $\cpij$ are usually noisy. 
Therefore, a second step is needed to ensure one can still obtain 
the approximate solution in the presence of measurement errors.
Zhang17 chose to minimize the difference between $\cpij$ and the product of $P^TC^R(\theta)P$,
\begin{equation}
\label{eq:minJ}
\text{min }{\cal J}\left(P;C^R(\theta)\right) \equiv \frac{1}{2}\sum_\theta \norm{\cpij -P^TC^R(\theta)P}_F^2
\end{equation}
where $\norm{...}_F^2$is the Frobenius norm.
The minimization is not trivial given the constraints in Eq.~\ref{eq:constraints}.
Since all the matrices in Eq.~\ref{eq:Cgg} and ~\ref{eq:constraints} should be non-negative \footnote{There could exist negative values in the measurements of $\cpij$, depending on the data quality.}, Zhang17 proposed the second algorithm in light of non-negative matrix factorization (NMF; \citealt{NMF1999}) method. They derived the corresponding ``multiplicative update rules''\footnote{See the derivation of update rules in the Appendix of Zhang17.} to 
minimize the the objective ${\cal J}$ iteratively and at the same time, meet the constraints of Eq.~\ref{eq:constraints}.
Given the difference between
the standard bifactor NMF (eg., min $\norm{V -WH}_F^2$, $V, W, H \in \mathbb{R}^{m\times n}$, $\mathbb{R}^{m\times r}$, $\mathbb{R}^{r\times n}$) and 
the trifactor NMF here (Eq.~\ref{eq:minJ}), the derived update rules guarantee a non-increasing
objective ${\cal J}$ when the initial input of $P$ is in the vicinity of the true
$P$, which is provided by the results from the first step. 
The self-calibration algorithm has been tested against the mock galaxy catalogue and found a good recovery in the scattering matrix $P$ at the level of 0.01-1 per cent.

During our application to observation, we modify
the algorithm to better meet our needs. 
The modifications are pretty minor: 
\begin{itemize}
\item we require the initial guess of scattering matrix $P$ being a random
diagonal-dominant matrix\footnote{The matrix A is diagonally dominate if $ |a_{ii}| \ge \sum_{j\neq i} |a_{ij}|$ for all $i$, where $a_{ii}$ denotes the elements in the $i$th row and $j$th column.}; 
\item we stop the iterations when ${\cal J}$ increases significantly. 
\end{itemize}
We find the first modification necessary thanks to the non-singular 
nature of the diagonal-dominant matrix. Otherwise, singular matrices can be encountered during
iterations. The second modification is also motivated by avoiding algorithm crashes.
Note that these two modifications are identical to those adopted in the
companion paper \citet{Peng2022}, who have tested on the mock catalogues
and found a less than 0.03 difference in the reconstructed scattering matrix $P$, 
compared with the ground truth (see their Table 1).

\begin{table}
\centering
\caption{
The cpu-time (in seconds) used by {\sc numba}-compiled self-calibration code in various configurations of photo-$z$ bins and input correlations.
Note that, the recorded time is employing a conservative max iteration number in
both fixed-point algorithm ($N_{\rm max} = 1000$) and NMF algorithm ($N_{\rm max} = 10000$). Compared to the traditional Markov Chain Monte Carlo approach, 
the self-calibration code is extremely efficient in arriving at the best solutions 
(e.g., 95 free parameters in the scenario of $N_\theta$ = 15 angular bins and 5 photo-$z$ bins). The self-calibration code is publicly available at \url{https://github.com/alanxuhaojie/self_calibration}. 
}\vspace*{1.5mm}
\begin{tabular}{|l|c|c|c|c|}
\hline
\diagbox[height=3\line]{photo-$z$ \\ bins \#}{angular \\ bins \#} & $N_\theta$ = 15 & $N_\theta$ = 10 & 
$N_\theta$ = 5 & $N_\theta$ = 2\\\hline
5 photo-$z$ bins  & 3.7 & 2.5 & 1.4 & 0.7 \\\hline
10 photo-$z$ bins & 6.6 & 4.3 & 2.3 & 1.3 \\\hline
\end{tabular}
\label{table:performance}
\end{table}

Our main difference lies in choosing the best scattering matrix $P$
among iterations and estimating its associated uncertainty. 
Without accounting for errors in measurements 
$C_{ij}^{P}(\theta)$, \citet{Peng2022} 
estimated the mean scattering matrix 
$P$ by averaging over the scattering matrices with 
low ${\cal J}$ values and
its uncertainty as the standard deviation in those matrices.
The uncertainty estimated this way effectively reflects the 
randomness in the algorithm, i.e., different initial guesses may
end randomly in the vicinity of truth. This approximation is
valid only when the measurements are accurate and precise.
To propagate the measurement uncertainty to scattering matrix $P$, 
we instead draw $N$ realizations of ``angular 2PCFs measurements'', 
assuming that the uncertainty of measured angular 2PCFs follows a 
Gaussian distribution. When drawing the realizations, 
the covariance between measurements has been accounted for. 
In addition, we pick up the scattering matrix $P$ 
with the lowest $\chi^2$ value among the iterations for all realizations.
The $\chi^2$ is defined as the sum of $\chi^2_{ij}$ of $i$th and $j$th
photo-z bin\footnote{By the definition of Eq.~\ref{eq:chi2}, we implicitly ignore the covariance among the 
measurements from different sets of auto- and cross-correlation function, 
which is unrealistic to obtain given small number of jackknife sub-samples.},
\begin{equation}
\chi^{2}=\sum_{ij}\chi^2_{ij} = \sum_{ij} \left(C^{P}_{ij}-C^{P,*}_{ij}\right)^{\mathrm{T}} \mathbf{Cov}_{ij}^{-1}\left(C^{P}_{ij}-C^{P,*}_{ij}\right)
\label{eq:chi2}
\end{equation}
where $C^{P,*}_{ij} \equiv \sum_{k}P_{ki}P_{kj}C^R_{kk}(\theta)$. 
$\mathbf{Cov}$ is the covariance matrix
estimated through 200 jackknife subsamples (cf. section ~\ref{subsec:clustering}). 
We note that after averaging over many realizations, the median Med($P$) 
and standard deviation NMAD($P$) 
remain almost the same if we choose the $P$ with the lowest ${\cal J}$ value among
iterations. See appendix~\ref{app:chi2_iterations} for a zoom-in
view how $\chi^2$ and $\cal{J}$ vary as a function of iterations. 

To be less affected by outliers, we assess
the mean and standard deviation of $N$ scattering matrices from $N$ realizations 
by adopting median and normalized median absolute deviation, 
dubbed Med($P$) and NMAD($P$), respectively.
The NMAD is defined as $1.48 \times$ median $|P-\text{median}(P)|$,
which converges to standard deviation for an ideal Gaussian distribution.
The median and NMAD are estimated elementwise, 
which may cause a slight deviation of sum-to-unity in columns of Med($P$).
Through out the paper, all presented 
matrices of Med($P$) and NMAD($P$) are estimated using 
$N=100$ realizations, and in each realization, we choose 
the scattering matrix $P$ with the lowest $\chi^2$ value among iterations.
We have verified that both Med($P$) and NMAD($P$) 
remain almost identical when using $N=200$. 
We summarize the methodology in Fig.~\ref{fig:flowchart}.

Since the self-calibration algorithm uses {\sc numpy} arrays and loops heavily, we accelerate
the {\sc python} code by {\sc Numba}\footnote{\url{https://numba.pydata.org/}}.
The performance of {\sc numba}-compiled code for one data realization is listed in Table~\ref{table:performance}.

\section{Results}
\label{sec:results}

\begin{figure*}
\includegraphics[width=\textwidth]{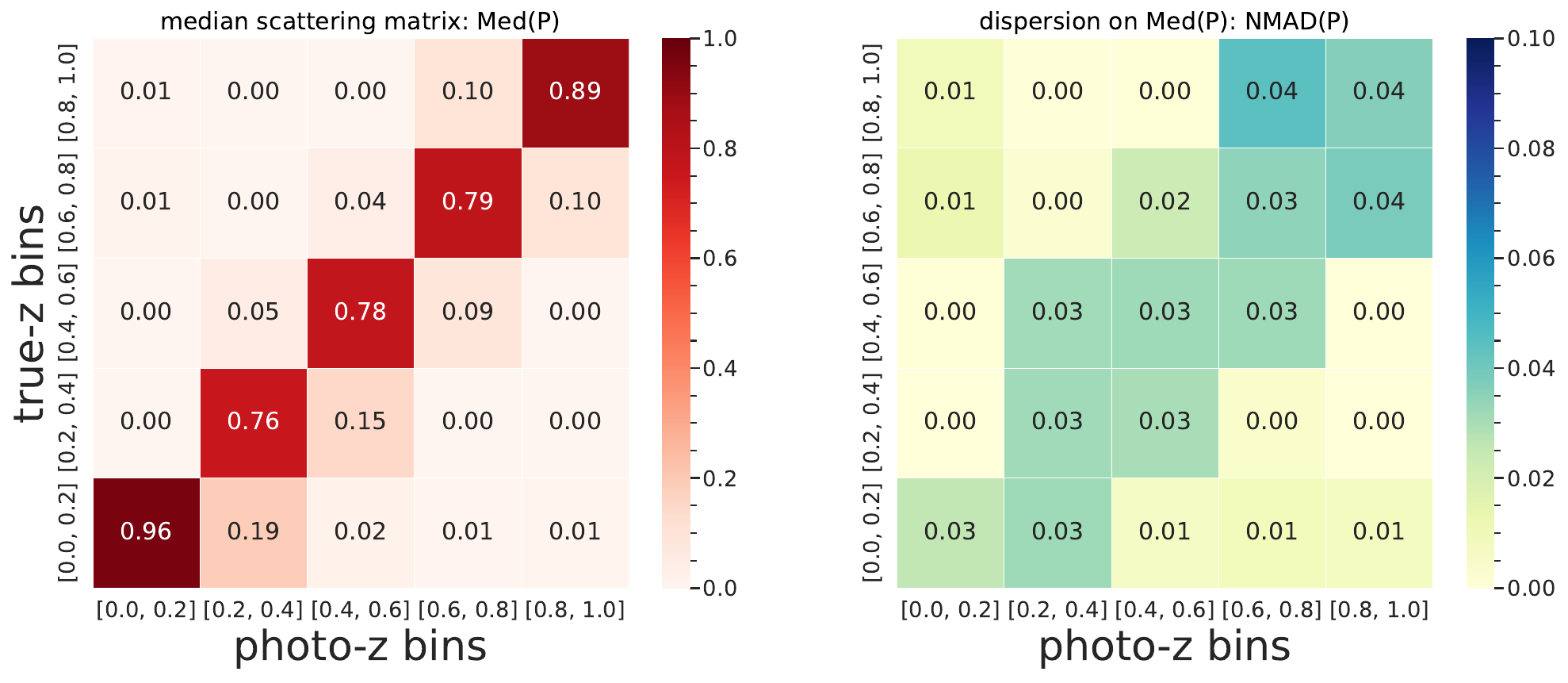}
\caption{
The median scattering matrix Med($P$) and normalized
median absolute deviation NMAD($P$) estimated from 
100 realizations of angular 2PCFs measurements 
for 5 tomographic bins.
In both panels, columns (rows) represent photometric (spectroscopic) 
redshift bins.
{\it Left}: the entry $P_{ij}$ stands for
the percentile of galaxies in $j$th photo-$z$ bin that is from $i$th spectroscopic-$z$ bin. 
For example, the bottom left entry $P_{12} = 0.19$ means that
19\% of galaxies in the 2nd photo-$z$ bin $0.2<z_{\rm photo}<0.4$ 
are actually from the 1st redshift bin $0<z_{\rm spec}<0.2$.
The majority of leaks are happening between neighbouring redshift bins.
No strong signal is detected for catastrophic photo-$z$ outliers.
Note that, $\sum_i$ Med($P_{ij}$) $\simeq 1$, where 
approximation comes from the fact that we do the median entrywise.
{\it Right}: The normalized median absolute deviation.
}
\label{fig:pij_5_redshift_bins}
\end{figure*}

We present the main results in section~\ref{subsec:fiducial}. 
Cosmic magnification induces correlations between tomographic bins, 
and we discuss its impact on the scattering matrix in section~\ref{subsec:mag}. 
As a consistency check, in section~\ref{subsec:10_redshift_bins}
we apply our algorithm to 10 tomographic bins
with $\Delta z = 0.1$ in the same redshift range and check whether
the scattering matrix for 10 bins agrees with that for 5 bins.
In section~\ref{subsec:sgc}, 
we apply our algorithm to DECaLS-SGC to see if the scattering
matrix agrees with the fiducial one, which also serves as a sanity check
for residual imaging systematics.
As long as Eq.~\ref{eq:Cgg} is solvable, 
the scattering matrix from the self-calibration algorithm
should be {\it independent} on angular scales of input correlations.
We test this expectation in section~\ref{subsec:scale}.
In the last section~\ref{subsec:KNN}, we compare the fiducial
scattering matrix with those from the power spectrum and weighted
spectroscopic subsample that approximates the photo-$z$ 
performance of the entire photometric sample.

\begin{figure*}
\includegraphics[width=\textwidth]{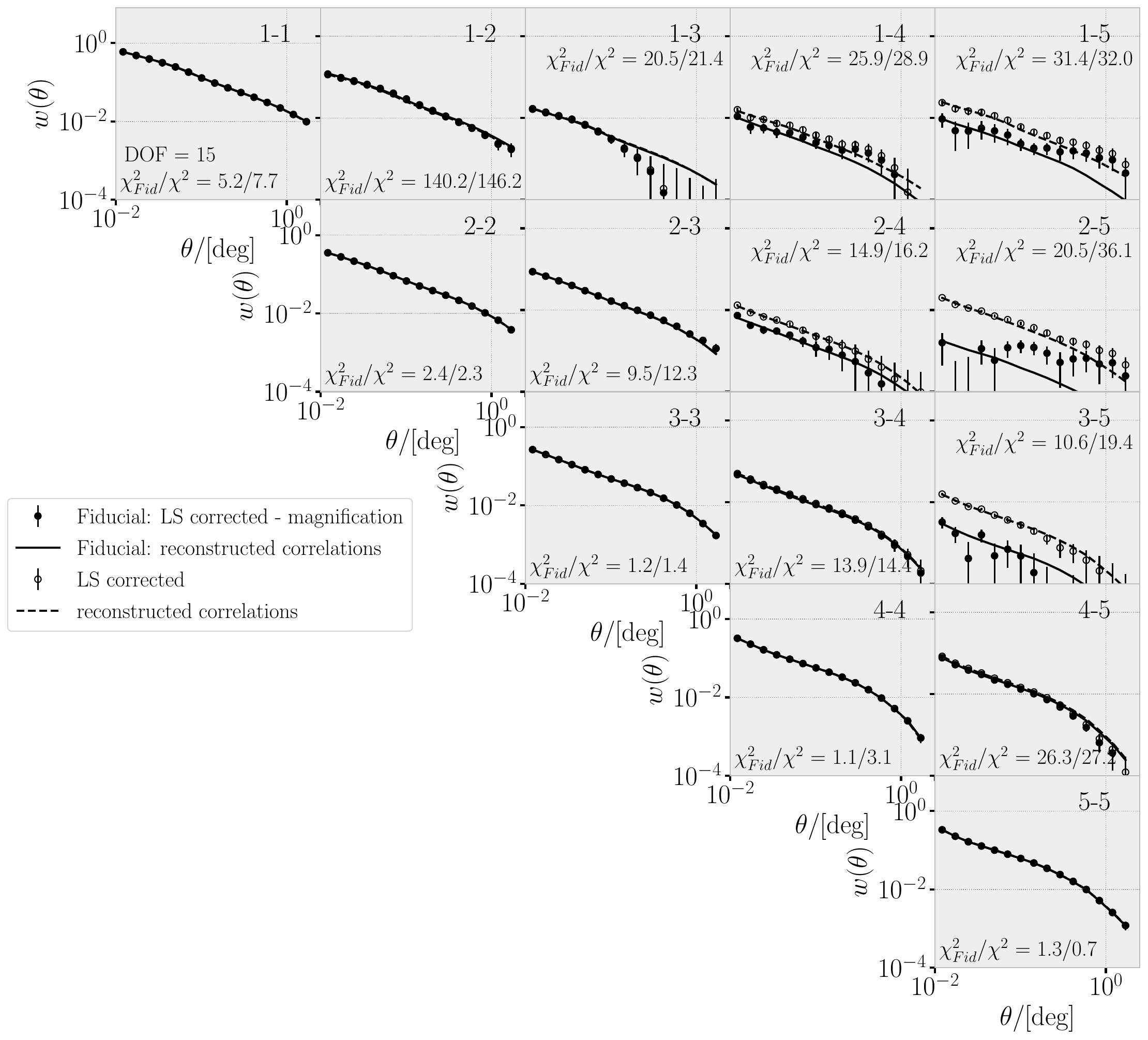}
\caption{
Two sets of correlation measurements versus their corresponding reconstructed correlations.
Both sets of measurements have been corrected for imaging systematics. The difference lies
in that, the fiducial ones (filled circles) have deducted the approximated correlations induced by cosmic magnification (solid red lines in Fig.~\ref{fig:mag}, section~\ref{subsec:mag}) from the systematics corrected measurements (empty circles, the same as red lines in Fig.~\ref{fig:2pcf}).
The lines are the median of 100 realizations of reconstructed correlations $\sum_{k}P_{ki}P_{kj}C^R_{kk}(\theta)$, within each realization we
pick up the scattering matrix $P$ with the lowest $\chi^2$ (Eq.~\ref{eq:chi2}) among iterations. 
The solid (dashed) lines are obtained from inputting filled circles (empty circles) as $C^{P}_{ij}$ in Eq.~\ref{eq:chi2}. 
Each panel represents either auto- or cross-correlations, whose tomographic bin indices are indicated at the top right corner. The corresponding $\chi^2$ are also listed for two scenarios. The covariance matrices are obtained through 200 jackknife subsamples.
}
\label{fig:dataVSmodel_5_redshift_bins}
\end{figure*}

\subsection{Fiducial results}
\label{subsec:fiducial}

\begin{figure*}
\includegraphics[width=\textwidth]{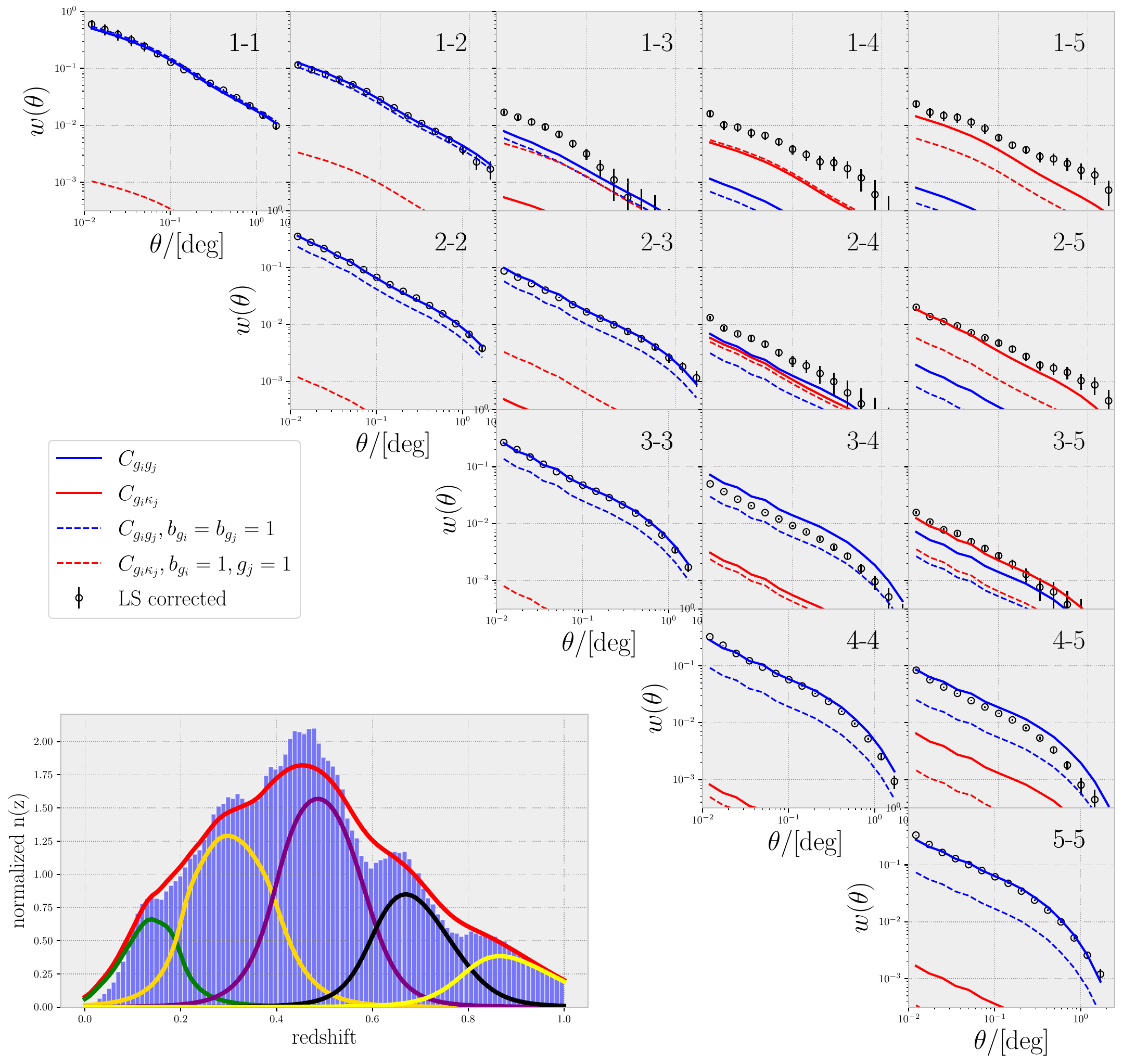}
\caption{
The induced correlations by cosmic magnification for 5 tomographic bins. 
{\it Insert panel}: The filled histograms show the photometric redshift
distribution of the default samples provided by the PRLS catalogue. 
The solid lines are stacked 
distributions by convolving photometric redshift with its error.
The red line is for the entire sample and the rest five lines for 5 tomographic bins.
{\it Corner plot}: The empty circles are the measured correlations that have been corrected for imaging systematics. Given the PRLS redshift distributions, 
the blue dashed lines show the intrinsic clustering assuming unity linear bias. 
The blue solid lines are the product of blue dashed lines with
linear bias factor $b_{g_i}b_{g_j}$, where $b_g$ = [0.96, 1.25, 1.38, 1.75, 1.94], 
and indices $ij$ are indicated at the top right corner in each panel.
The numerical values of $b_g$ are obtained the square root of the ratio of 
empty circles and blue dashed lines in diagonal panels. The red solid lines
represent the cosmic magnification $b_{g_i}g^j\langle\delta_{m}^{i}\kappa^{j}\rangle$, where
$g$ = [-1.47, -1.21, 0.12, 0.94, 2.54]. For reference, 
we also plot the cosmic magnification 
without applying the linear bias and $g$ factor scaling. 
All lines are calculated via package {\sc ccl}
\citep{cclpaper}, assuming the Planck 2018 cosmology \citep{Planck2018}.
}
\label{fig:mag}
\end{figure*}

After accounting for the cosmic magnification (see 
the following subsection for details), we show
the median scattering matrix Med($P$) and 
its normalized median absolute deviation NMAD($P$) in Fig.~\ref{fig:pij_5_redshift_bins}. 
The scattering matrix suggests that $\sim 80\%$ of galaxies remain in their redshift bin, except the first and the last redshift bin. 
The majority of scattering happens between neighbouring redshift bins. 
The right-hand panel of Fig.~\ref{fig:pij_5_redshift_bins} reflects the 
fluctuation of elements in the scattering matrices in different realizations, 
a typical fluctuation $\sim 0.03$ suggesting high confidence in the values of median scattering matrix. In addition, we do not find a clear signal of photo-$z$ outliers or catastrophic 
photo-$z$ errors in fiducial galaxy samples.

The comparison between measured ($\cpij$) and reconstructed angular 
2PCFs ($P^TC^R(\theta)P$) is shown in Fig.~\ref{fig:dataVSmodel_5_redshift_bins}. 
These two agree well in autocorrelations and 
neighbouring-bins cross-correlations at the full scales. 
The large $\chi^2$ value in panel 1-2 is due to the noisy covariance
matrix, and if only the diagonals are used, the corresponding 
$\chi^2$ would decrease to $\sim 14$.
The good agreements are even achieved for cross-correlations in 
neighbour-next-neighbour bins 
(e.g., panels 2-4 and 3-5), except panel 1-3 
where the measurements display a sudden drop at large scales. 
The reconstructed correlations seem to only disagree with the measurements 
when the shape of measurements departs from a power law. 
Though the current self-calibration algorithm is not optimal in
terms of $\chi^2$\footnote{The
update rules are by design to minimize objective $\cal{J}$ , 
rather than $\chi^2$. 
Therefore, the returned scattering matrix would be more determined by those higher numerical values of measurements for their higher weights in $\cal{J}$. 
One can expect that, in parts of higher measurement values, the agreements between measured and reconstructed 2PCFs are better, as seen in Fig.~\ref{fig:dataVSmodel_5_redshift_bins}. 
An optimal way should take into account measurement uncertainties when iterating for the best solution.}, 
the agreement presented in Fig.~\ref{fig:dataVSmodel_5_redshift_bins} is surprisingly good. 
After all, the algorithm is to mathematically decompose the observed correlations matrix into three component matrices and no astrophysical or cosmology parameters are assumed. 
The reconstructed 2PCFs even display some hints for
1- and 2-halo term transition.

\begin{figure*}
\includegraphics[width=\textwidth]{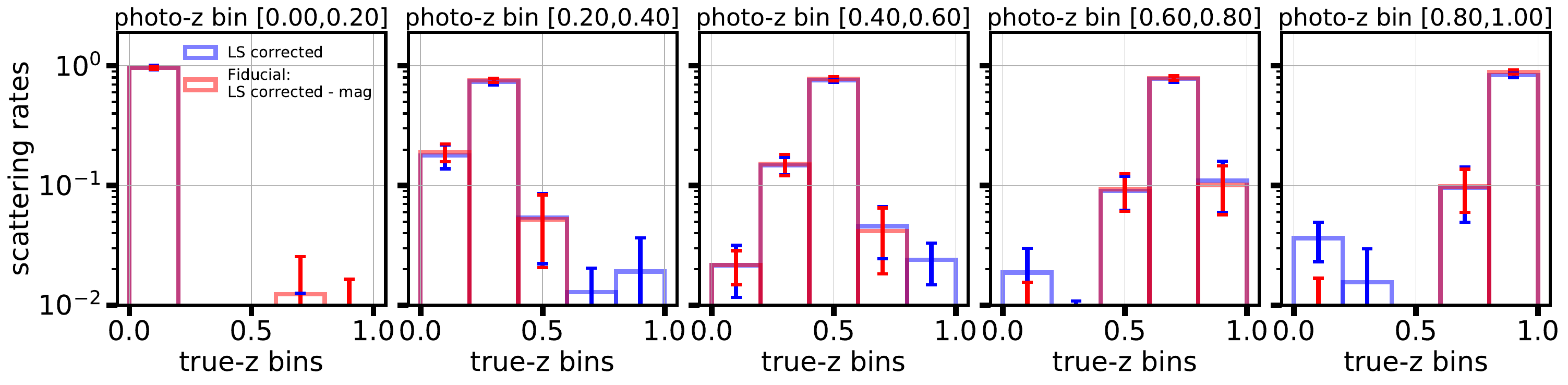}
\caption{
Comparison of median scattering matrix Med($P$) 
with and without removal of cosmic magnification. 
Each panel represents a column of median scattering matrix shown in Fig.~\ref{fig:pij_5_redshift_bins}, which describes the underlying
redshift distribution in a given photo-$z$ bin. 
The red bars are the fiducial results that have removed the cosmic magnification. They present the same information as in Fig.~\ref{fig:pij_5_redshift_bins}, 
where the heights and error bars are the corresponding entries of
median scattering matrix and normalized median absolute deviation, respectively.
The blue bars are the results without removal of cosmic magnification.
Note that, the measured cross-correlations lead to $\sim$ 2-4 per cent photo-z outlier, if without removal of cosmic magnification.
}
\label{fig:mag_P}
\end{figure*}

\subsection{Cosmic magnification}
\label{subsec:mag}
The foreground dark matter distribution changes the path of light
from the background galaxies. The so-called gravitational lensing
magnifies the surface area per solid angle so that (1)
it dilutes the galaxy surface densities, and (2) it makes the galaxies appear 
brighter because lensing conserves the surface brightness. 
If the galaxy luminosity function above flux $f$ follows a power law 
$n(>f) \propto f^{-\alpha} $, then in the weaklensing regime 
the observed galaxy overdensity $\delta_{g}^{o}$ can be approximated as
\begin{equation}
\delta_{g}^{o}=\delta_{g}+ g\kappa
\end{equation}
where $\delta_{g}$ is the true galaxy overdensity and 
$g \equiv 2(\alpha-1)$. The $\kappa$ is the lensing convergence, depicting
the lensing effect caused by matter in front of the galaxies.
The second term on the right hand side is introduced by cosmic magnification. 
When cross-correlating two tomograpic bins,
the actual measurements are (see also e.g., \citealt{Moessner1998}),
\begin{equation}
\langle\delta_{g}^{o,f}\delta_{g}^{o,b}\rangle=
\langle\delta_{g}^{f}\delta_{g}^{b}\rangle + 
g^b\langle\delta_{g}^{f}\kappa^{b}\rangle +
g^f\langle\kappa^{f}\delta_g^{b}\rangle +
g^fg^b\langle\kappa^{f}\kappa^{b}\rangle 
\label{eq:mag}
\end{equation}
where $\delta_{g}^{o,f}$ and $\delta_{g}^{o,b}$ are 
observed galaxy overdensity in foreground and
background bin, respectively.
On the right-hand side, 
the first term is the intrinsic clustering 
when two tomographic bins overlap in redshift. 
The second (third) term describes the lensing of background
(foreground) galaxies by the front matter traced by foreground 
(background) galaxies.
The last term reflects the lensing of background and foreground galaxies
by the common dark matter in front of them. 
The Eq.~\ref{eq:mag} tells that there will be some correlations 
even when two tomographic bins do not overlap in redshift. 
It breaks our assumption that the cross-correlations are contributed 
{\it only}
by the redshift overlaps. Since 
the last two terms are usually tiny compared to the second term,
we will ignore these two terms and focus on the second
term, $g^b\langle\delta_{g}^{f}\kappa^{b}\rangle$.
We note that dust propelled by star formation activity and AGN 
in foreground galaxies will also induce correlations
between foreground and background galaxies \citep{Menard2010, Fang2011}.
We ignore this effect for simplification.

Before feeding to the self-calibration algorithm, 
one should remove the cosmic magnification contribution from the measured 2PCFs.
However, it is not a trivial task to accurately estimate the cosmic
magnification term $g^b\langle\delta_{g}^{f}\kappa^{b}\rangle$. 
We approximate the induced 2PCFs by cosmic magnification (cf. Fig.~\ref{fig:mag}) 
by (1) adopting the fiducial redshift distribution provided in the PRLS catalogue (insert panel in Fig.~\ref{fig:mag}); 
(2) estimating the linear galaxy bias of tomographic bins
by the square root of the ratio of measured
autocorrelations (black filled circles in diagonal panels) and theory prediction (blue dashed lines in diagonal panels);
(3) measuring luminosity slope at flux limit $z_{\rm mag} = 21$ in each tomographic bin. 
From the diagonal panels in Fig.~\ref{fig:mag}, the simple linear
bias assumption seems to work quite well at interested angular scales. The agreement also
demonstrates that 
the fiducial redshift distribution is a good approximation of the underlying redshift distribution. 
For far-away
cross-correlations, the contributions from cosmic magnification are significantly higher
than those from intrinsic clustering, i.e., order-of-magnitude higher in panel 1-5. 
It would detect a fake signal of photo-$z$ outlier if 
mistakenly interpreting the measured correlations contributed 
by photo-$z$ overlap. We also find that the shape of correlations from magnification 
is very similar to that of intrinsic clustering, which raises some difficulties 
in separating them.
However, we note that the exact contribution by 
magnification is highly uncertain, depending
on our knowledge about galaxy bias, luminosity slope, and the underlying redshift
distribution. None of them can be trivially obtained.  
One should treat our present analysis as a zero-order approximation.

To see the impact of cosmic magnification on the scattering matrix, 
we feed two sets of correlation measurements into the self-calibration algorithm. 
Both sets have been corrected for imaging systematics. 
One set is
the correlation measurements without dealing with cosmic magnification
(empty circles in Fig.~\ref{fig:dataVSmodel_5_redshift_bins} and ~\ref{fig:mag}). 
Another set is the one with removing the approximated cosmic magnification contribution 
(filled circles in Fig.~\ref{fig:dataVSmodel_5_redshift_bins}). 
We compare the median scattering matrix in Fig.~\ref{fig:mag_P}. 
The cosmic magnification seems to have little impact on leading entries 
(diagonal and first-off-diagonal elements). 
However, without removal of cosmic 
magnification, the measured cross-correlations lead to
$\sim2-4$ per cent photo-$z$ outlier. 

\subsection{Finer photo-$z$ bins: 10 redshift bins}
\label{subsec:10_redshift_bins}
It is interesting to see if our algorithm can handle a much larger input matrix.
For $N_\theta = 15$ angular bins, the number of free parameters goes from 95 
(5 tomographic bins) to 240 (10 tomographic bins).
This 2.5-fold increase of free parameters will pose a challenge to the stability of
the self-calibration algorithm. Also, as a sanity check, we would like to test whether 
the finer scattering matrix from 10 tomographic bins agree with that from 5 bins.

\begin{figure*}
\includegraphics[width=\textwidth]{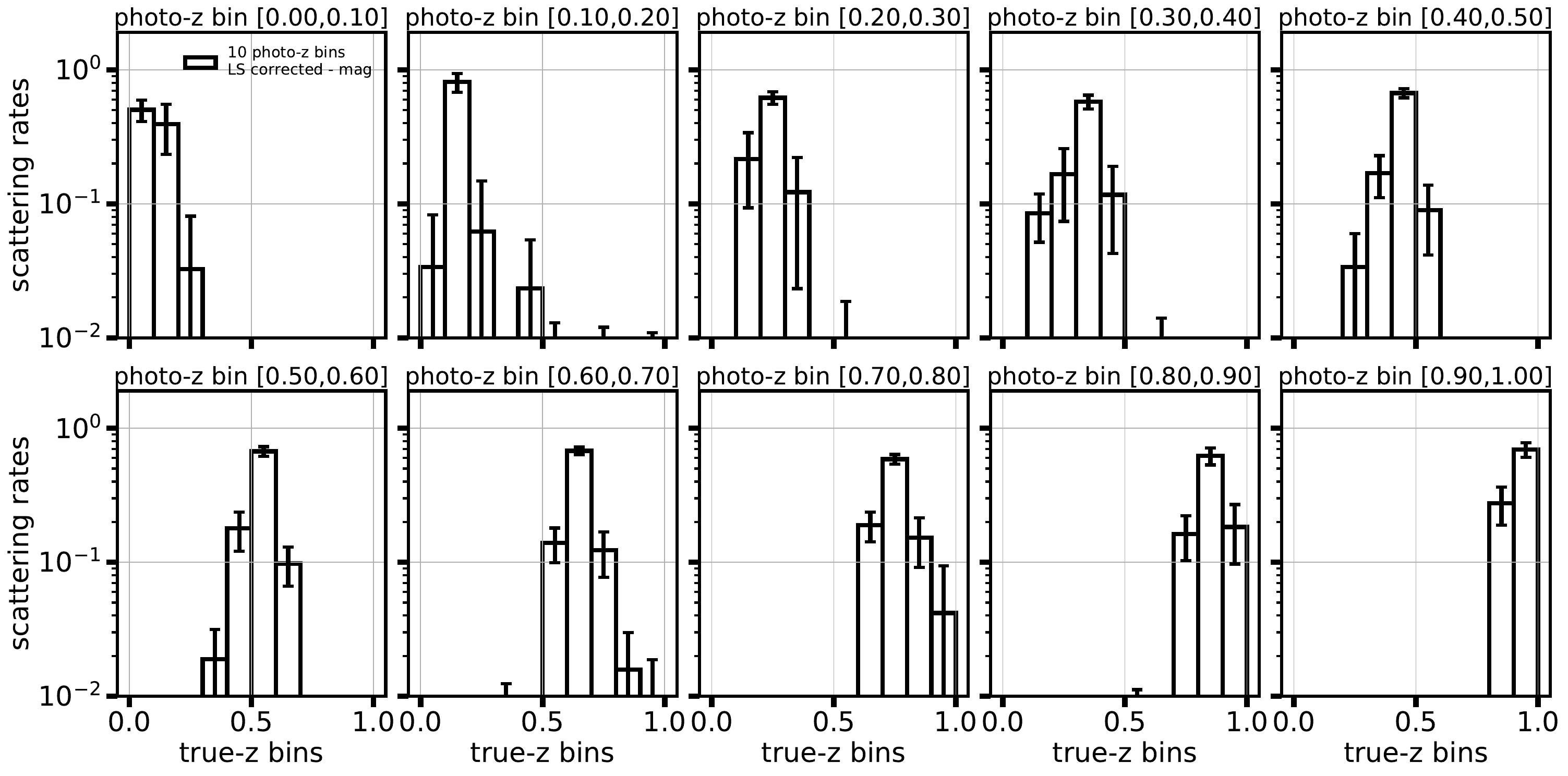}
\caption{The median scattering matrix from 10 tomographic bins.
The scattering matrix looks reasonable with each panel displaying 
a high peak in its own redshift bin and a more or less symmetric
wing extending to neighbouring bins. The reasonable scattering matrix
proves that the self-calibration code also works for the 
scenarios with 10 redshift bins. 
Again, we do not find a strong signal for catastrophic photo-$z$ outliers.
}
\label{fig:mag_P10}
\end{figure*}

We bin the fiducial sample into 10 tomographic bins with $\Delta z = 0.1$. 
We correct the imaging systematics for each finer tomographic bin 
and measure the auto- and cross-correlations. 
We also estimate the cosmic magnification contribution and deduct them from the measurements.
The resultant scattering matrix is presented in Fig.~\ref{fig:mag_P10}, 
which looks reasonable with a high peak at its
redshift bin and a wing on both sides.
Again, we find little evidence for photo-$z$ outliers 
after removing the cosmic magnification.

\begin{figure*}
\includegraphics[width=\textwidth]{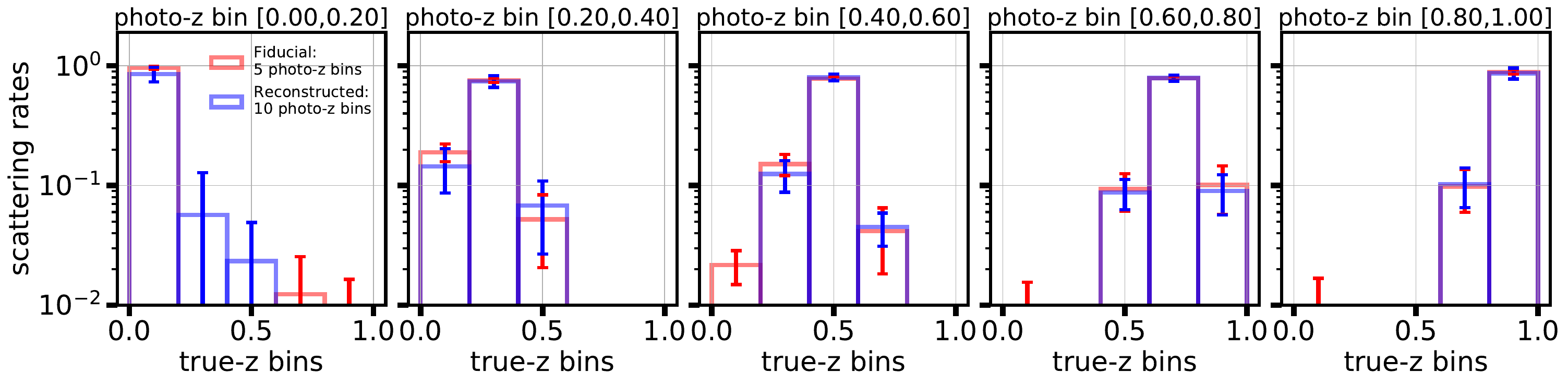}
\caption{Comparison of the fiducial median scattering matrix (red bars) and the 
scattering matrix of 5 photo-$z$ bins (blue bars) reconstructed from 10 photo-$z$ bins (Fig.~\ref{fig:mag_P10}). 
The error bars on reconstructed scattering matrix 
are estimated through error propagation 
from the 10 photo-$z$ bins scattering matrix. During the
propagation, we ignore the correlations between uncertainties
in the high-resolution matrix.
}
\label{fig:P10_P5}
\end{figure*}

With the number of galaxies in tomographic bins, 
it is straightforward to reconstruct a 
low-resolution scattering matrix (5 tomographic bins) 
from a high-resolution scattering matrix (10 tomographic bins). 
Figure ~\ref{fig:P10_P5} shows the comparison between
the fiducial and reconstructed scattering matrices.
The two matrices are broadly consistent with each other,
though the reconstructed matrix suggests considerable scattering rates
in the first tomographic bin.
To summary, our algorithm works for 10 tomographic bins and 
the agreement with the fiducial matrix are encouraging. 
We have also verified that our algorithm works for 20 tomographic bins.

\subsection{Scale dependence of scattering matrix?}
\label{subsec:scale}
\begin{figure*}
\includegraphics[width=\textwidth]{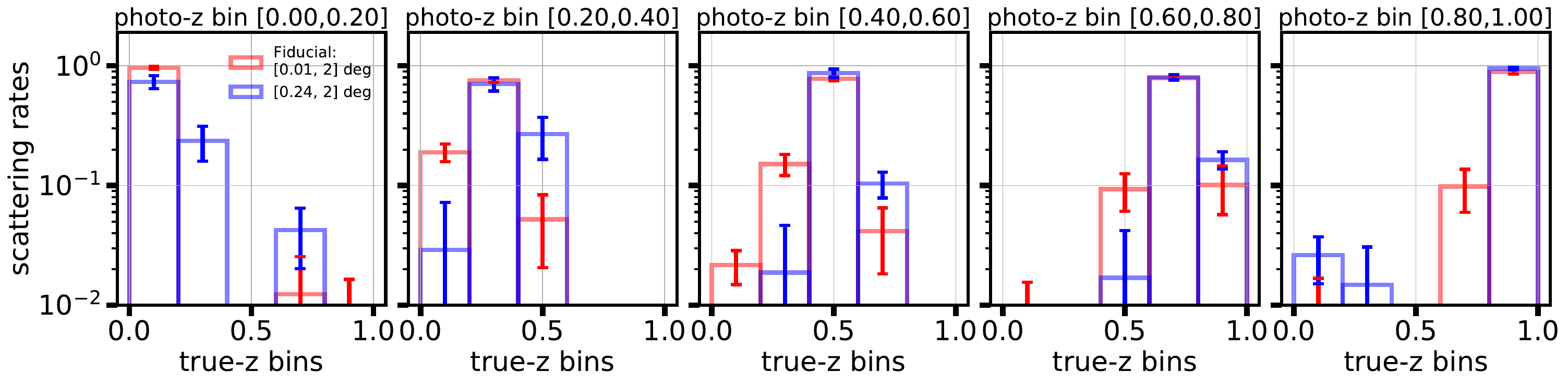}
\caption{Scale dependence of scattering matrix. The legend in the leftmost panel indicates
the angular scale cut of input correlations.
Red bars are the fiducial results, while 
blue bars represent the scattering matrix obtained 
with correlation function within the angular range
$[0.25, 2]$ deg, above which the imaging systematics are largely mitigated. 
}
\label{fig:scale_dependence_P}
\end{figure*}

Another interesting test is to see whether the solution 
scattering matrix depends on the angular scales of input correlations.
We solve for the scattering matrix by feeding the self-calibration code
with angular
correlations above $\sim 0.23$ deg (right to vertical lines in Fig.~\ref{fig:2pcf}).
The comparison is presented in Fig.~\ref{fig:scale_dependence_P}.
We notice some tension between these two matrices.
For example, the fiducial results suggest that 
$\sim 95\%$ of $0<z_{\rm photo}<0.2$ galaxies remain in their redshift bin, 
while the scattering matrix from large-scale correlations suggests $\sim 75\%$.
Since each column in the scattering matrix is subject to the sum-to-unity, 
$\sim 20\%$ of galaxies in the first photo-$z$ bin, 
suggested by the scattering matrix from large-scale correlations, comes from second
true-$z$ bin. Using clustering information only, 
it is difficult to break the degeneracy between the scattering
$i \rightarrow j$ and $j \rightarrow i$. 
Therefore, to fit the observed $C^{P}_{12}$, the fiducial
matrix estimates that $\sim 20\%$ of galaxies in the second
photo-$z$ bin come from the first true-$z$ bin.
This degeneracy could probably explain the tension
shown in Fig.~\ref{fig:scale_dependence_P}. 

\begin{figure*}
\includegraphics[width=\textwidth]{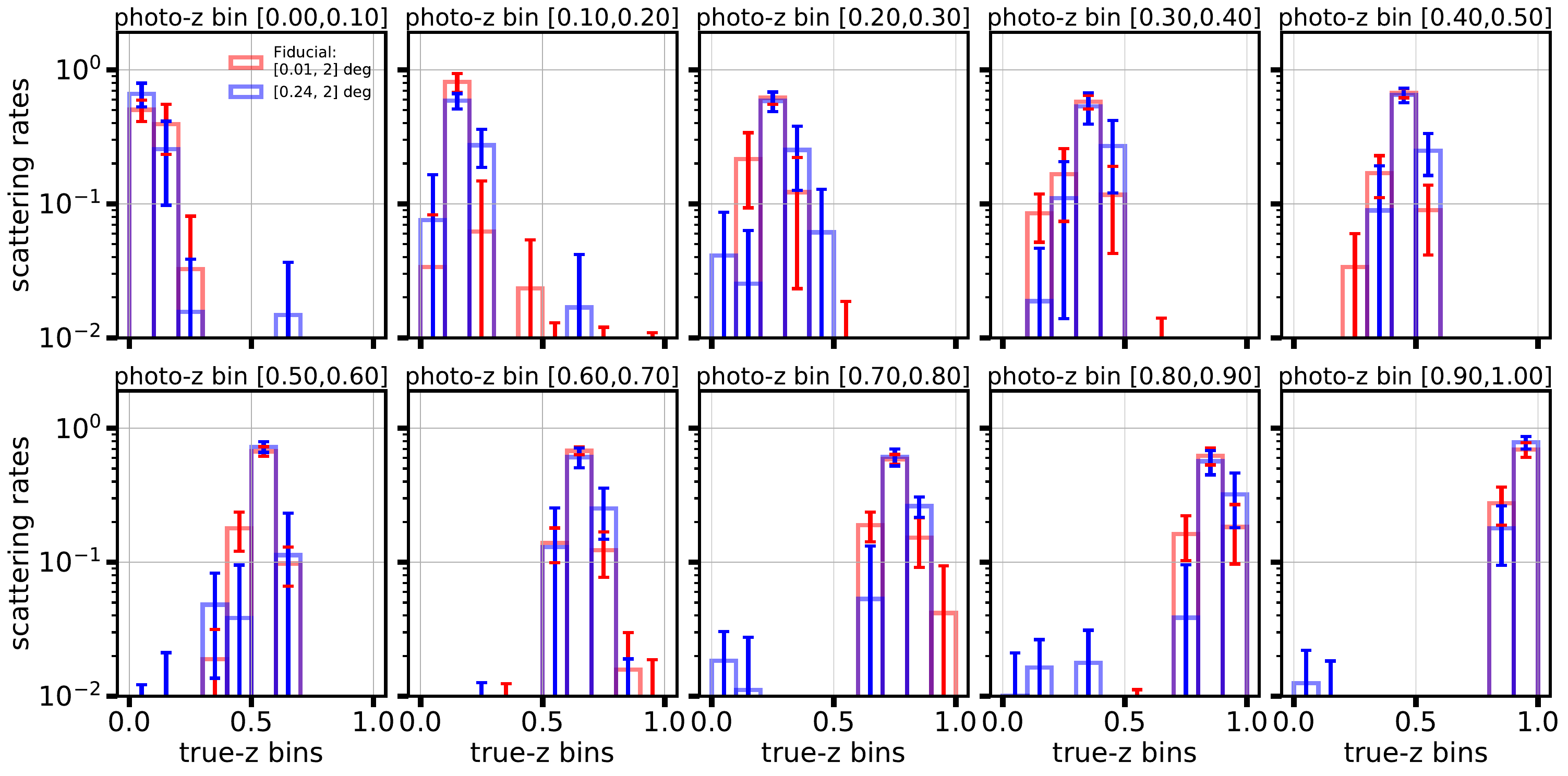}
\caption{Scale dependence of scattering matrix for 10 tomographic bins. 
}
\label{fig:scale_dependence_P10}
\end{figure*}

As shown in Figure 4 of \citet{Benjamin2010}, finer tomographic bins help 
alleviate the degeneracy since narrower bin width is
more sensitive to the photo-$z$ errors. 
Figure~\ref{fig:scale_dependence_P10} shows the scale dependence
of scatter matrix for 10 photo-$z$ bins. Overall, 
the agreement is much better, though error bars are larger and 
some degeneracy still exists. 

The scale-dependent tension may be caused by 
the incorrect imaging systematics correction used in this work.
For example, \citet{Rodriguez-Monroy2022} points out that machine-learning
based correction may lead to an oversuppression of clustering signals, 
which would not appear if using a classical linear approach.
To what extent the systematics correction would affect the scattering matrices,
we feed the uncorrected correlations (blue symbols or lines in Fig.\ref{fig:2pcf}) 
into the self-calibration algorithm. For a given scale range, 
the returned scattering matrices are almost the same, 
before and after the systematics correction.
That is to say, the scale-dependent tension persists, with or without 
systematics correction.
The existence of such tension in these two extreme scenarios suggests that  
it is more likely related to the intrinsic degeneracy aforementioned.

\subsection{DECaLS-SGC}
\label{subsec:sgc}

\begin{figure*}
\includegraphics[width=\textwidth]{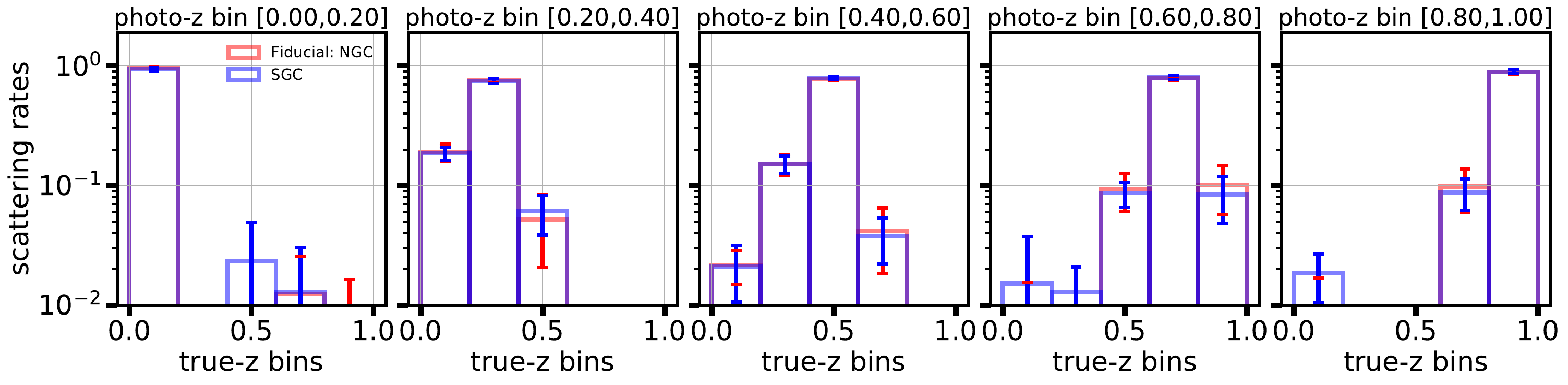}
\caption{Comparison of scattering matrix from fiducial sample
and DECaLS-SGC. Within the errors, they are almost identical to each other, which also confirms that the 
residual imaging systematics is minimal for all samples in both regions.
}
\label{fig:SGC_NGC}
\end{figure*}

\begin{figure*}
\includegraphics[width=\textwidth]{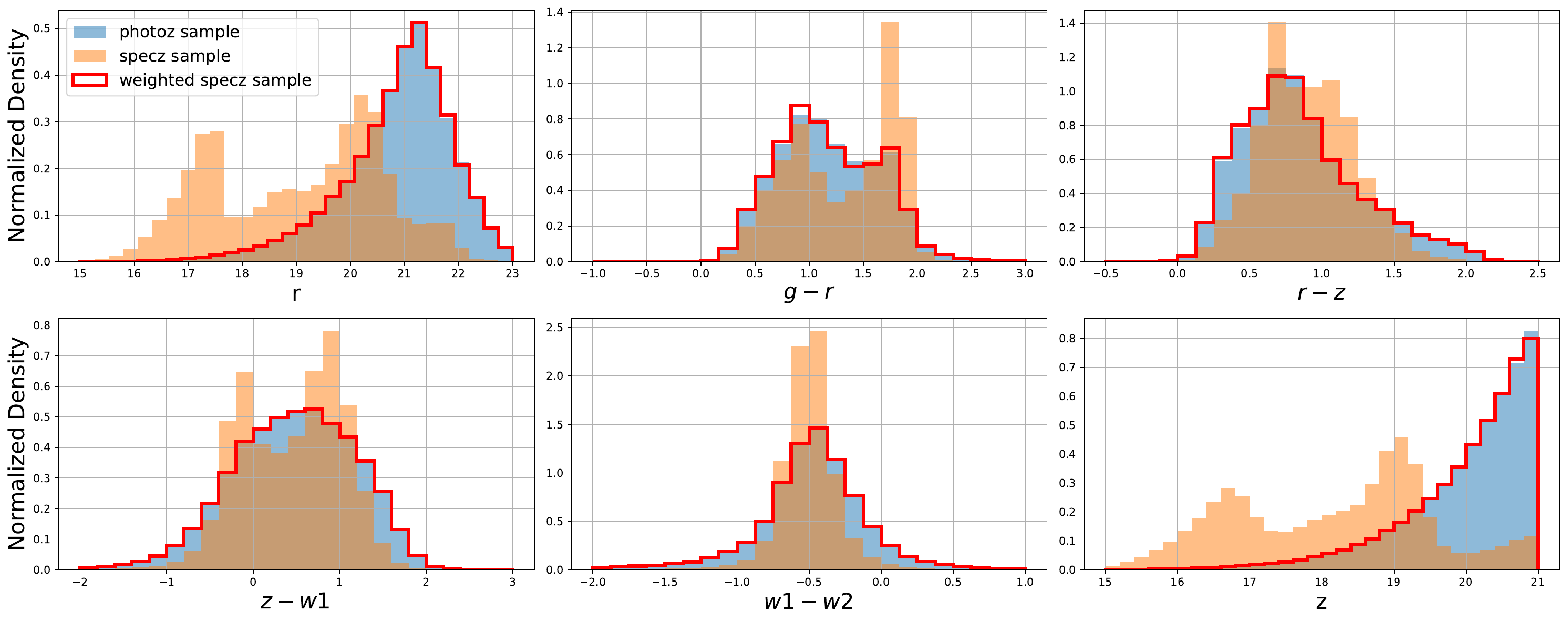}
\caption{The galaxy distributions in magnitude--colour space of photometric sample (light blue filled histograms), spectroscopic subsample (light orange filled histograms), and weighted spectroscopic subsample (red steps). Each panel compares the histograms of three samples
in magnitudes or colors. We have normalized the distributions for comparison. 
After weighting, the weighted spectroscopic subsample matches the photometric sample in magnitude--colour space by design.
}
\label{fig:mag_color_space_reweight}
\end{figure*}

\begin{figure*}
\includegraphics[width=\textwidth]{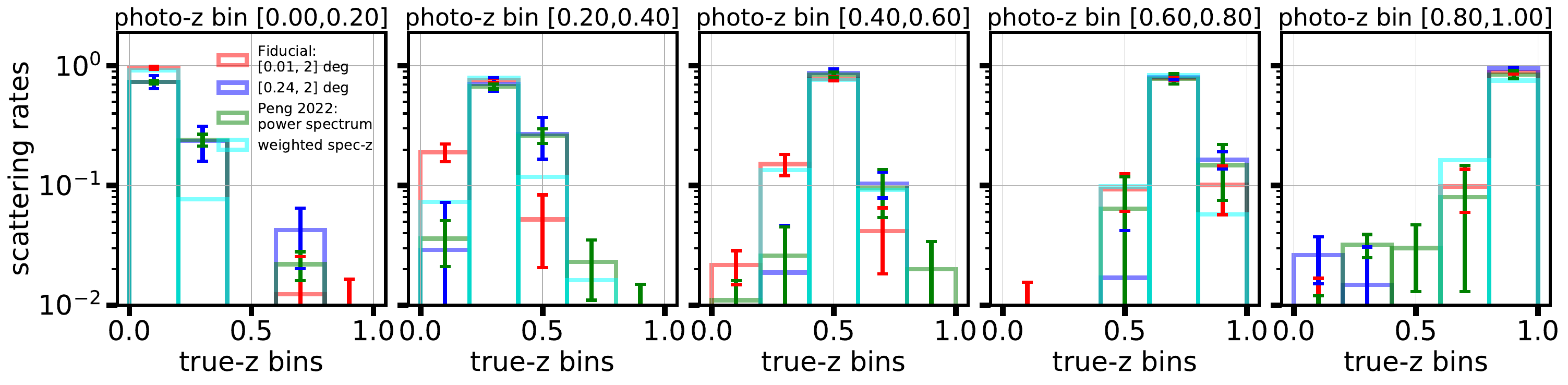}
\caption{Comparison of scattering matrices from external methods. 
The red and blue bars are the fiducial scattering matrix and the one obtained 
with correlation function in the angular range $[0.25, 2]$ deg, respectively.
The green bars are adopted from their Figure.10 in \citet{Peng2022}, 
by applying the self-calibration algorithm to the power spectrum measured
from DECaLS Data Release 8. The cyan bars are calculated from
the weighted spectroscopic subsample, which by design matches the
colour--magnitude distribution of the fiducial sample.
 Note that the four matrices should be seen as approximations to the truth, and these approximations suffer from different systematics (see reasons in section~\ref{subsec:KNN}). However, the overall agreements between approximations demonstrate that they should be reasonable, at least in the first order.
}
\label{fig:external_P}
\end{figure*}

We apply the same sample selections,
random catalogue construction, and imaging systematics mitigation
to DECaLS-SGC (see Appendix~\ref{app:sgc} for figures). 
We also estimate the cosmic magnification and
deduct it from correlation measurements. 
The scattering matrix is shown in Fig.~\ref{fig:SGC_NGC},  
consistent with the fiducial matrix.
The agreement suggests that the redshift distributions are similar, 
and that the residual imaging systematics is minimal.
We have verified that the same conclusions hold for 10 redshift bins.

\subsection{External comparisons}
\label{subsec:KNN}
In previous subsections, we conduct several self-consistency checks
regarding cosmic magnification, finer photo-$z$ bin width, 
scale dependence, and different sky coverage. In this subsection,
we would like to compare the scattering matrix with those estimated from 
power spectrum and weighted spectroscopic subsample.

Our companion paper \citet{Peng2022} applied the self-calibration 
algorithm to power spectrum and obtained the scattering matrix.
The power spectrum is measured from the same footprint as
the fiducial sample in this work, but from the DECaLS Data Release 8, which
is almost identical. 
The slight difference in algorithm has been discussed in section~\ref{sec:method}.
Figure~\ref{fig:external_P} shows the scattering matrices comparison. 
The scattering matrix from power spectrum is quite close to that 
from large scale correlation function. 

To facilitate another external test, 
we compare our scattering matrix with that estimated from 
weighted spectroscopic subsample, which approximates the photo-$z$ performance of
the entire photometric sample \citep{Lima2008, Bonnett2016}. Figure~\ref{fig:mag_color_space_reweight} shows that the spectroscopic subsample
is biased to brighter magnitudes and the colour distribution is not representative to that of photometric sample.
We could alleviate this difference by upper (or down) weighting the spectroscopic galaxies.
We cross match the photometric sample
to the spectroscopic subsample in multidimensional 
magnitude-colour space of $r$-band magnitude, 
$g-r$, $r-z$, $z-W1$ and $W1-W2$ colours (the magnitude-colour information used to train the colour-redshift relation, see also section 3.5 of \citealt{Zhou2020}). 
Each photometric galaxy is linked to its nearest spectroscopic neighbour in magnitude-colour space. The weight for a spectroscopic galaxy is therefore the number counts of photometric galaxies it linked to. 
This procedure guarantees that the weighted
spectroscopic subsample matches the photometric sample in terms of galaxy distribution in magnitude-colour space, which is shown in Fig.~\ref{fig:mag_color_space_reweight}.
The weighted spectroscopic subsample can provide a more fair assessment of photo-$z$ performance. 

From the weighted spectroscopic subsample, we calculate the scattering matrix by definition. 
The comparison is
shown in Fig.~\ref{fig:external_P}. 
Two scattering matrices overall agree with each other.

 We emphasize that the point of Fig.~\ref{fig:external_P} is to check to what extent the overall shapes of the four scattering matrices agree, rather than the exact percentages they differ by. The reason is that all four matrices shown in Fig.~\ref{fig:external_P} should be seen as rough estimations of the true scattering matrix, and it is OK that approximations differ. However, these approximations suffer different systematics. For example, the three scattering matrices returned from the self-calibration method may suffer from scale-dependent tension, galaxy distribution bias (see definition in the second to the last paragraph in section~\ref{sec:dis}), and the notorious degeneracy between up and down scattering rates. The one from the weighted spectroscopic subsample is evaluated by the closest neighbour in multi-dimension colour--magnitude space, i.e., k=1 in kNN. Determining the exact number of neighbours to use that gives an unbiased matrix is not a trivial task, which can be investigated in galaxy mocks. The overall agreements between approximations demonstrate that they should be reasonable, at least in the first order.



\section{Summary and Discussion}
\label{sec:dis}

Inaccurate photo-$z$ introduces cross-correlations between different photo-$z$
bins and the amplitude of correlations 
is proportional to scattering rates.
The idea that uses a set of auto- and cross-correlations to constrain the redshift distribution of photometric samples has long
been explored from the theoretical perspective \citep{Schneider2006, Zhang2010}. 

\citet{Benjamin2010}
solved the scattering rates by assuming
that the cross-correlations between two photo-$z$ bins 
come from the scattering between these very two bins. 
In reality, any common contamination from a third redshift bin  
would also induce correlations.
Therefore, their simplification may bias the inferred 
redshift distribution, which
may not meet the redshift accuracy requirement for the 
LSST-like projects. 
\citet{Zhang2017} developed the self-calibration algorithm, 
which is able to obtain the
exact solution for scattering 
rates for ideal mock data. 

In this work, we implement the self-calibration algorithm to observational photometric
galaxy survey, the DECaLS Data Release 9. 
We improve the algorithm by starting with a more reasonable initial
guess and by adjusting the convergence criterion. 
These two improvements greatly enhance the stability of the 
algorithm when facing with noisy measurements. In addition,
we select the scattering matrix with the lowest $\chi^2$ value, 
rather than the ${\cal J}$ value (cf. Eq.~\ref{eq:minJ} and Fig.~\ref{fig:chi2_iterations}) that the algorithm minimizes in, to be more physically quantified.  
Finally, we propagate the measurement uncertainties to the final scattering 
matrix by drawing
realizations of measurements assuming that 
the measurements follow Gaussian
distribution (see details in section~\ref{sec:method}).

On the observation side, we correct for the spurious correlations due to various observational conditions,
including but not limited to Galactic extinction, seeing, and stellar density (19 imaging maps in total).
We employ a machine learning method to mitigate the imaging systematics.
We mitigate the imaging systematics for each interested tomography sample to make sure
the auto- and cross-correlations contain the minimal contamination.
Please see the details in section~\ref{subsec:imaging}.

With the improved algorithm and 
decontaminated correlation measurements, we list the main findings as following: 

\begin{itemize}

\item The self-calibration algorithm works for angular correlations measured
from observational photometric catalogue with 5 equal-width redshift bins in $0<z_{\rm photo}<1$. 
The majority ($\sim 80$ per cent) galaxies stay in their own redshift bin (cf. Fig.~\ref{fig:pij_5_redshift_bins}).
Most leaks happen between neighbouring redshift bins.
We do not see a strong signal for photo-$z$ outlier in fiducial samples. 

\item Cosmic magnification induces correlations between photo-$z$ bins, which may bias the scattering matrix if not properly accounted for. 
We approximate the magnification shown in Fig.~\ref{fig:mag}.
We compare the scattering matrices with and without accounting for magnification 
in Fig.~\ref{fig:mag_P}. It seems that the scattering matrix changes little
after accounting for magnification.
However, we emphasize that a few percent ($\sim 3\%$) 
galaxies will be mistakenly considered as
photo-$z$ outlier if the magnification is not accounted for.

\item The self-calibration algorithm also works for 10 photo-$z$ bins, 
in which free parameters increase from 95 (5 bins) to 240 (10 bins). 
The scattering matrix from 10 photo-$z$ bins (cf. Fig.~\ref{fig:mag_P10})
renders a finer redshift distribution in photo-$z$ bins, which
suffers less degeneracy compared to wider bin width (see also figure 4 in
\citealt{Benjamin2010}).

\item The scattering matrix from 10 photo-$z$ bins can be easily 
downgraded to one of 5 bins, which serves a sanity
check when compared to the fiducial scattering matrix (cf. Fig.~\ref{fig:P10_P5}).
The two matrices show some tension that might be attributed to
the degeneracy of scattering rates.

\item The self-calibration algorithm applies to both non-linear
and linear scales. We compare the scattering matrix from
relative large scales to the fiducial one (cf. Fig.~\ref{fig:scale_dependence_P}).
In principle, we expect two matrices to agree with each other but 
some tension shows instead. Again, we suspect that it might
relate with degeneracy in scattering rates, supported by 
the agreement being much better for 10 photo-$z$ bin scenario (cf. Fig.~\ref{fig:scale_dependence_P10}).

\item The scattering matrix from the South Galactic Cap
is almost identical with the fiducial one (cf. Fig.~\ref{fig:SGC_NGC}),
which further confirms that residual imaging systematics is minimal after
mitigation.

\item Comparison of scattering matrices constructed 
from external methods serves a strong test for the self-calibration algorithm. 
We compare the scattering matrices from power spectrum and 
weighted spectroscopic subsample. The overall agreement (cf. Fig.~\ref{fig:external_P}) demonstrates the feasibility of
the self-calibration algorithm, though some level of tensions
exists.
\end{itemize}

Although some tension shows up between various scattering matrices, 
it is encouraging that the self-calibration algorithm works for
the noisy observational measurements and that these scattering matrices
agree with each other reasonably. 
The self-calibration method
provides an alternative way to calibrate the redshift distribution
for photometric samples. We emphasize that the method does
not rely on any cosmological prior nor parametrization of 
photo-$z$ probability distribution, which is particularly helpful in 
constraining the 
equation state parameters of dark energy 
in a typical 3$\times$2 analysis \citep{Schaan2020}.

We note that there is an important implicit assumption 
in the self-calibration method.
In Eq.~\ref{eq:Cgg}, 
we assume that galaxies scatter from $i$th true-$z$ bin to $j$th photo-$z$ bin
share the same intrinsic clustering with those who remain,
i.e., $C^R_{ii}(\theta) = C^R_{ij,i\neq j}(\theta)$. 
In reality, that might not be this case. For example, 
galaxies that scatter further are probably overall fainter 
because faint galaxies 
are typically less sampled by the spectroscopic sample. 
In addition, it is well known that galaxy bias varies as a function
of luminosity and colour (eg. \citealt{Zehavi2011, Xu2018c, Wang2021}).
Ignoring such difference would lead to a problematic scattering matrix. 
One future work is to evaluate to what extent the variation of galaxies bias 
could affect the accuracy of scattering matrix. 
In addition, we assume that there are no galaxies scattering to or from redshift bin
$z > 1$ since we restrict our analysis to galaxy sample with $z_{\rm photo} < 1$. 
However, this assumption
is also probably problematic because there should exist a considerable fraction of
galaxies scattering to or from redshift beyond our redshift range, especially
for the highest redshift bin. For instance, all photo-$z$ bins, except the first bin, show a 
comparable scattering rates to their neighbouring bins. This pattern is 
expected in the highest redshift bin, which, however, is prohibited in our analysis.
Since all scattering rates are correlated, it is hard to tell how much such 
simplification
would affect the final scattering matrix. It may also contribute to the tension between the scattering matrices as presented in various tests. 

Using galaxy clustering alone cannot distinguish the upper and down scatter, e.g. fig.11 in \citealt{Erben2009} and fig.2 in \citealt{Benjamin2010}. However, the upper and down scatter would leave a distinct
galaxy--galaxy lensing signal. Therefore, including galaxy--galaxy lensing information 
would greatly break this degeneracy \citep{Zhang2010}. 
We reserve this investigation for future work. Another direction that could 
advance the self-calibration method is to optimize the algorithm.
As we mentioned in section~\ref{sec:method}, current algorithm is to minimize 
the objective $\cal{J}$, 
square difference between data and model (cf. Eq.\ref{eq:minJ}).
It probably makes the algorithm to more likely return a solution that is preferred
by larger numerical values of input measurements, e.g., angular correlations, 
power spectrum or galaxy--galaxy lensing. 
An optimal way should take into account measurement uncertainty
when iterating for the best solution.

\section*{Acknowledgements}
HX would like to thank Edmond Chaussidon and Mehdi Rezaie 
for their detailed answer to the questions on imaging systematics mitigation.
The authors thank the referee for helpful comments.
This work is supported by the National Key R\&D Program of China (2018YFA0404504, 2018YFA0404601, 2020YFC2201600), National Science Foundation of China (11833005, 11890692, 11890691, 11621303, 11653003), the Ministry of Science and Technology of China (2020SKA0110100), the China Manned Space Project with NO.CMS-CSST-2021 (B01 \& A02), the 111 project No. B20019, the CAS Interdisciplinary Innovation Team (JCTD-2019-05), and the Shanghai Natural Science Foundation (19ZR1466800) 

\section*{Data availability}
We make the self-calibration code publicly available at \url{https://github.com/alanxuhaojie/self_calibration}.
To avoid clutter, we showcase imaging systematics 
correction related figures only for the first 
tomographic sample $0< z < 0.2$ in the DECaLS-NGC region, e.g., Fig.~\ref{fig:systematics} and Fig.~\ref{fig:systematics_cross}. 
All figures for other tomographic samples are available on request.

\bibliography{ms}

\appendix
\section{$\cal J$ and $\chi^2$ as a function of
iterations}

\label{app:chi2_iterations}
\begin{figure*}
\includegraphics[width=\textwidth]{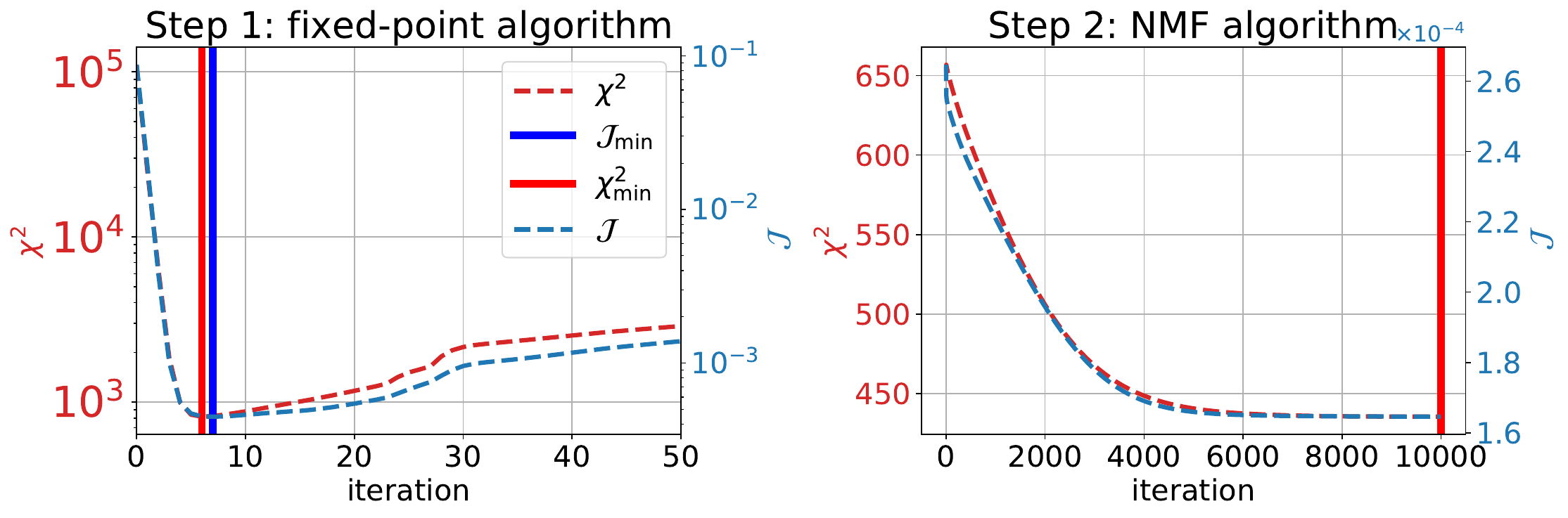}
\caption{
$\cal J$ and $\chi^2$ as a function of iterations in
the fixed-point algorithm (left) and NMF algorithm (right).
These two algorithms make up the self-calibration code.
The red (blue) dashed line shows the $\chi^2$ ($\cal J$)
as a function of iterations. The red (blue) vertical lines
marks the iteration where $\chi^2$ ($\cal J$) is the 
minimum. For each data realization, we choose the 
scattering matrix with lowest $\chi^2$ over the one with
the lowest $\cal J$, both of which are shown as vertical lines
in the right-hand panel. The plot is made with a random data realization
drawn from auto- and cross- correlations measured from the first tomographic sample $0< z < 0.2$ of the DECaLS-NGC region.
}
\label{fig:chi2_iterations}
\end{figure*}

In each data realization, we prefer the scattering
matrix with the lowest $\chi^2$ over the one
with lowest $\cal J$, which the self-calibration code
aims to minimize (see section~\ref{sec:method}).
This choice in fact changes little to the final
median scattering matrix Med($P$) and its uncertainty NMAD($P$) 
after averaging over many realizations (e.g., $N=100$ realizations used in this work).

In Fig.~\ref{fig:chi2_iterations}, we zoom into the
two child algorithms that make up the self-calibration code to see how $\cal J$ and $\chi^2$ change as codes iterate.
Within just a few iterations, the fixed-point algorithm finds a solution whose $\cal J$ (and $\chi^2$) drops almost two orders of magnitude. Then both $\cal J$ and $\chi^2$ start to increase slowly, which implies that the fixed-point algorithm can not reduce $\cal J$ (and $\chi^2$) anymore. Here comes the second step, the NMF algorithm. We start with
the scattering matrix with the lowest $\chi^2$ from the
first step. The NMF algorithm works to further reduce $\cal J$ (and $\chi^2$) by $\sim50\%$, before reaching a plateau.  Although the $\chi^2$--$\cal J$ relation is not exactly monotonic,  Fig.~\ref{fig:chi2_iterations} justifies  that
the self-calibration code also minimizes the $\chi^2$ effectively. 

\section{DECaLS South Galatic Cap}
\label{app:sgc}
The results presented in the main text are based on DECaLS-NGC (cf. Fig.~\ref{fig:footprint}). In this appendix, we showcase the footprint 
and imaging systematics mitigation 
of the DECaLS-SGC.

\begin{figure}
\includegraphics[width = \columnwidth]{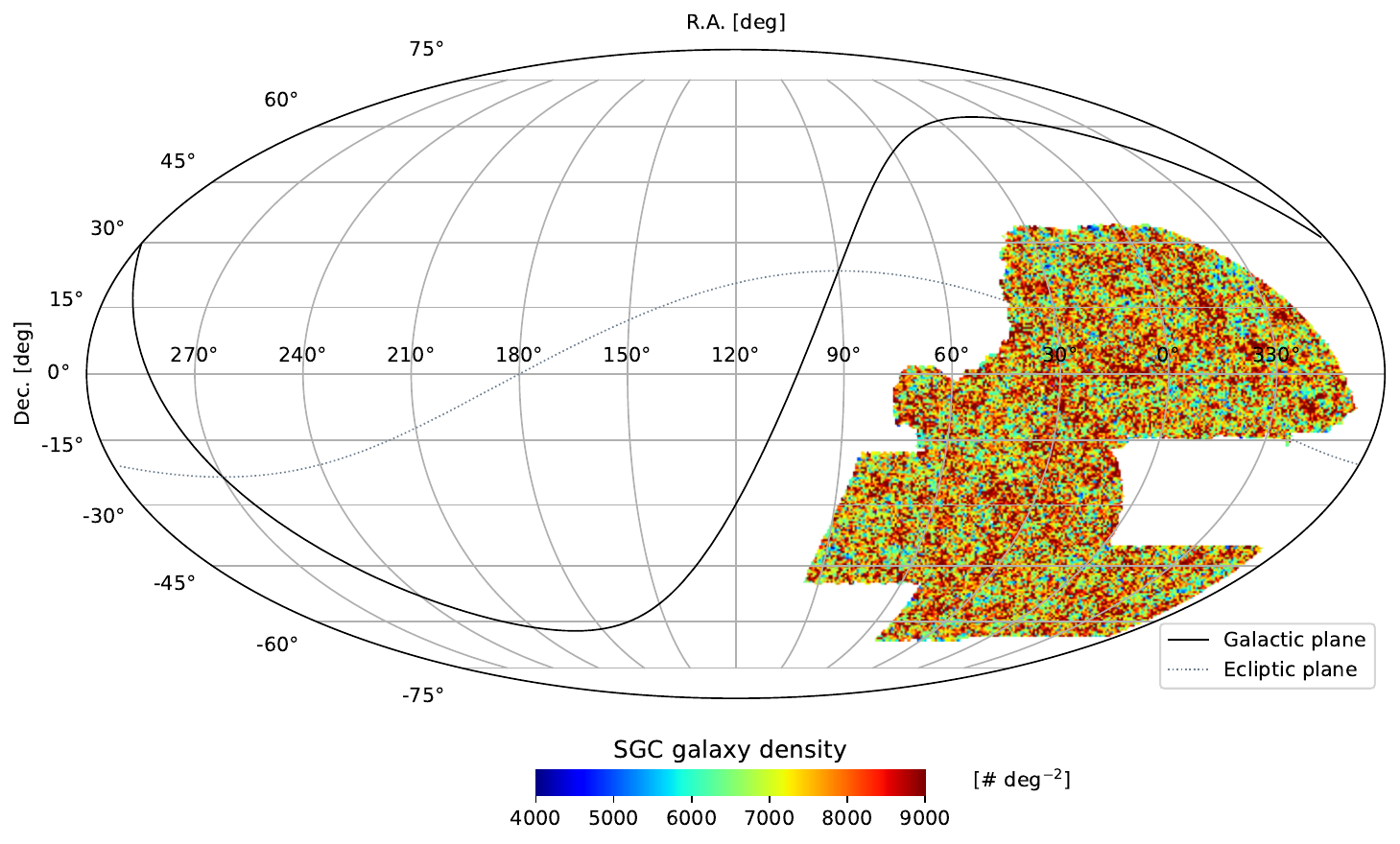}	
\caption{
The footprint of the DECaLS-SGC galaxy sample. 
The total population of galaxies is $\sim 59$ million with the sky
coverage $\sim 7637$ deg$^2$, a surface density $\sim 2.14$ galaxies
per square arcmin. The colour code is the galaxy number counts per deg$^2$. We note this figure does not take into account the 
fractional observed area. 
}
\label{fig:SGC_footprint}
\end{figure}

In addition to the sample selections in section~\ref{subsec:sample}, we also mask out the region close to the Large Magellanic Cloud. 
The final footprint of DECaLS-SGC is shown in Fig.~\ref{fig:SGC_footprint}.

\begin{figure*}
\includegraphics[width=\textwidth]{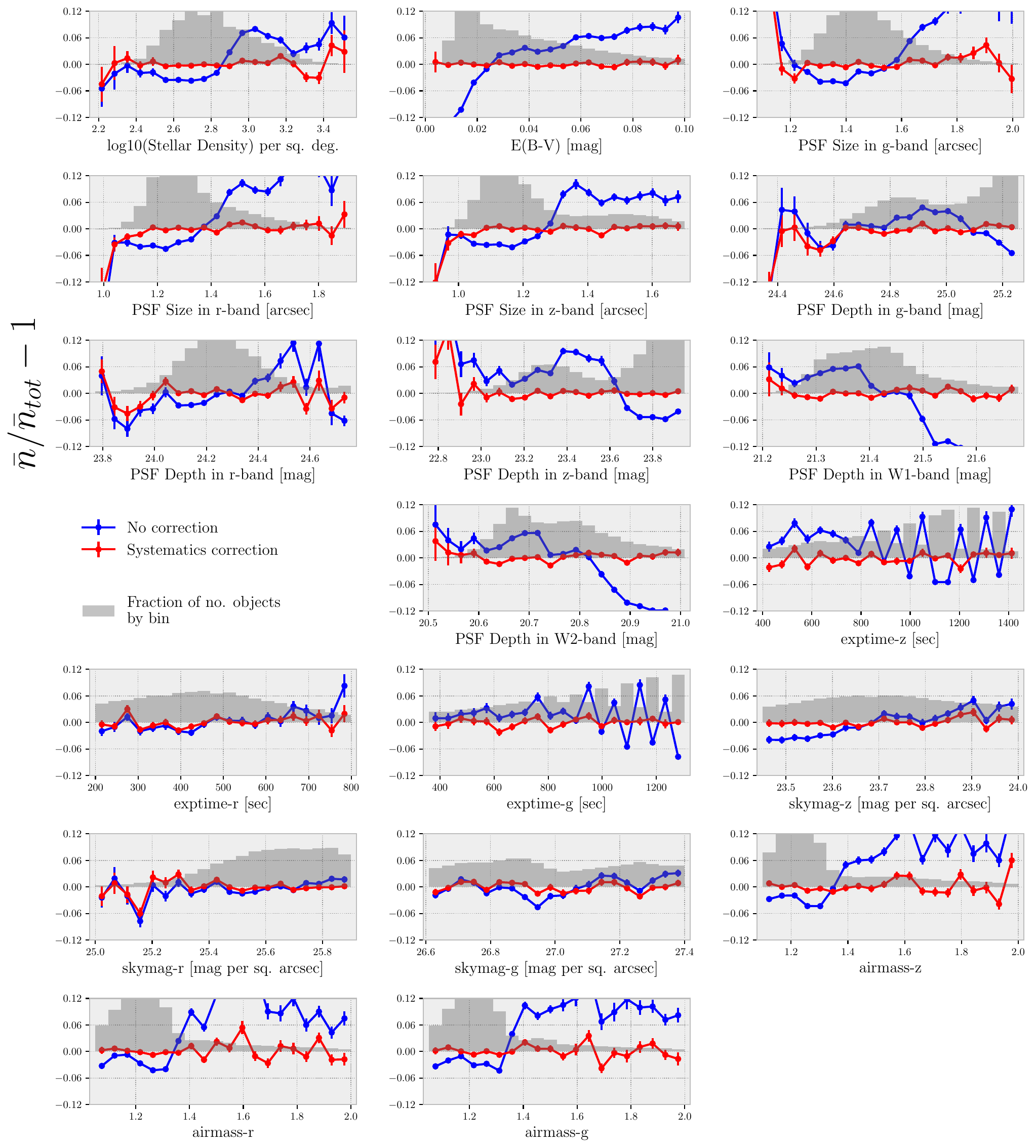}
\caption{
Relative galaxy overdensity as a function of 19 input imaging maps.
Similar to Fig.~\ref{fig:systematics} but for the $0 < z < 0.2$ sample in the DECaLS-SGC region.
The blue/red lines show the relative density before/after the systematics 
correction inside valid pixels. 
After the mitigation, the corrected density is much flatter 
versus all input imaging properties.
}
\label{fig:SGC_systematics}
\end{figure*}

We individually apply the imaging systematics mitigation procedure (cf. section~\ref{subsec:imaging}) to each tomographic bin in DECaLS-SGC. 
We showcase the galaxy density for $0 < z < 0.2$ sample, before and after the correction, as a function of various imaging maps 
in Fig.~\ref{fig:SGC_systematics}. The imaging systematics mitigation seems
to work pretty well (actually even better compared to the fiducial samples)
for DECaLS-SGC. 

\end{document}